\documentclass[sigconf,balance=false]{acmart}

\AtBeginDocument{%
  }

\usepackage{popets}
\setcopyright{popets}
\copyrightyear{YYYY}
\acmYear{YYYY}
\acmVolume{YYYY}
\acmNumber{X}
\acmDOI{XXXXXXX.XXXXXXX}
\acmISBN{}
\acmConference{Proceedings on Privacy Enhancing Technologies}
\usepackage[utf8]{inputenc}
\usepackage[table]{xcolor}
\usepackage{url}
\usepackage{xspace}
\usepackage{multirow}
\usepackage{pgfplots}
\usepackage{pgfplotstable}
\usepackage{booktabs}
\usepackage{eurosym}
\usepackage{makecell}
\usepackage{enumerate}
\usepackage{tabularx}
\usepackage{enumitem}
\usepackage{comment}
\usepackage[super]{nth}
\usepackage[labelfont=bf]{caption}

\PassOptionsToPackage{hyphens}{url}

\newenvironment{packeditemize}{\begin{list}{$\bullet$}{\setlength{\itemsep}{0.2pt}\addtolength{\labelwidth}{-4pt}\setlength{\leftmargin}{\labelwidth}\setlength{\listparindent}{\parindent}\setlength{\parsep}{1pt}\setlength{\topsep}{0pt}}}{\end{list}}

\newif\ifcomments
\commentsfalse

\definecolor{MyGreen}{rgb}{0.13,0.545,0.13}

\newcommand\ml[1]{\ifcomments\textcolor{magenta}{[[ML: #1]]}\fi}

\newcommand\nv[1]{\ifcomments\textcolor{red}{[[NV: #1]]}\fi}

\providecommand{\ie}{\emph{i.e.,\xspace{}}}
\providecommand{\eg}{\emph{e.g.,\xspace{}}}

\providecommand{\etal}{\emph{et al.\xspace{}}}
\providecommand{\etc}{\emph{etc.\xspace{}}}
\newcommand\para[1]{\noindent\textbf{#1}}

{}

\providecommand{\gateway}{\emph{Gateway}}
\providecommand{\director}{\emph{Director}}

\renewcommand{\texttt}[1]{%
 \begingroup
 \ttfamily
 \begingroup\lccode`~=`/\lowercase{\endgroup\def~}{/\discretionary{}{}{}}%
 \begingroup\lccode`~=`[\lowercase{\endgroup\def~}{[\discretionary{}{}{}}%
 \begingroup\lccode`~=`.\lowercase{\endgroup\def~}{.\discretionary{}{}{}}%
 \catcode`/=\active\catcode`[=\active\catcode`.=\active
 \scantokens{#1\noexpand}%
 \endgroup
}

\usepackage{relsize}

\usepackage{pifont} 
\newcommand{\dcircle}[1]{\ding{\numexpr201 + #1}}
\newcommand{\yes}{$\bullet$}
\newcommand{\no}{}

\def\appsUnder20{{4,200}}

\def\appsOver100{{63}}

\def\allOriginsMedianAOSPPerms{{13}}

\def\allOriginsMaxAOSPPerms{{101}}

\def\allOriginsMedianCustomPerms{{2}}

\def\allOriginsMaxCustomPerms{{130}}

\def\allOriginsMedianAllPerms{{15}}

\def\allOriginsMaxAllPerms{{231}}

\def\medianAOSPPermsAlt{{20}}

\def\maxAOSPPermsAlt{{101}}

\def\medianCustomPermsAlt{{2}}

\def\maxCustomPermsAlt{{130}}

\def\medianAllPermsAlt{{21}}

\def\maxAllPermsAlt{{231}}

\def\medianAOSPPermsGP{{11.5}}

\def\maxAOSPPermsGP{{39}}

\def\medianCustomPermsGP{{2}}

\def\maxCustomPermsGP{{79}}

\def\medianAllPermsGP{{14}}

\def\maxAllPermsGP{{106}}

\def\medianAOSPPermsPI{{17}}

\def\maxAOSPPermsPI{{32}}

\def\medianCustomPermsPI{{3}}

\def\maxCustomPermsPI{{8}}

\def\medianAllPermsPI{{20}}

\def\maxAllPermsPI{{40}}

\def\allRequestedPerms{{1,181}}

\def\nbCustomPerms{{1,075}}

\def\nbCustomPerms{{522}}

\def\priceBrowser{{26\%}}

\def\performanceBrowser{{44\%}}

\def\privacyBrowser{{42\%}}

\def\kidBrowser{{2\%}}

\def\totalBrowsers{{424}}

\def\loadedOneAuto{{381}}

\def\loadedOneManual{{19}}

\def\loadedOneTotal{{400}}

\def\loadedAllPages{{367}}

\def\onboardingBrowsers{{13}}

\def\privacyPolicyFirstBrowsers{{19}}

\def\noTest{{24}}

\def\leakPIIPercentage{{32\%}}

\def\totalBrowsersGPlay{{266}}

\def\totalBrowsersPre{{5}}

\def\totalBrowsersAlternate{{157}}

\def\onlyOneHash{{94\%}}

\def\maxCerts{{3}}

\def\playNonresettableRatio{{2\%}}

\def\playResettableRatio{{31\%}}

\def\altNonresettableRatio{{6\%}}

\def\altResettableRatio{{8\%}}

\def\perGoogleAds{{30\%}}

\def\moreOneMBrowsers{{56}}

\def\percGPlayBrowsers{{63\%}}

\def\wellKnownBrowsers{{266}}

\def\easylistBrowsers{{276}}

\def\easylistBrowsersPercentage{{65\%}}

\def\nonEasylistBlockersPercentage{{73\%}}

\def\analyticsBlockersPercentage{{60\%}}

\def\adBlockersPercentage{{88\%}}

\def\widgetBlockersPercentage{{77\%}}

\def\easylistButAllowed{{376}}

\def\easylistButAllowedPercentage{{89\%}}

\def\easylistPossiblyBlocked{{48}}

\def\easylistDefinitelyBlocked{{17}}

\def\easylistDefinitelyBlockedPercentage{{4\%}}

\def\trackerBlockers{{276}}

\def\percTrackerBlockers{{65\%}}

\def\userAgentWebViewBrowsers{{84}}

\def\webViewImplementers{{378}}

\def\percWebViewImplementers{{89\%}}

\def\webViewImplementerAndXRequested{{300}}

\def\percWebViewImplementerAndXRequested{{71\%}}

\def\xRequestedNotWebView{{17}}

\def\chromeBrowsers{{12}}

\def\percChromeBrowsers{{3\%}}

\def\firefoxBrowsers{{2}}

\def\percFirefoxBrowsers{{$<$1\%}}

\def\engineUnknownBrowsers{{110}}

\def\percEngineUnknownBrowsers{{26\%}}

\def\domainHttpsBrowsers{{10}}

\def\domainHttpsBrowsersPercentage{{2\%}}

\def\domainHttpsOnPlay{{1}}

\def\domainHttpsOnAlt{{9}}

\def\totalHTTPSFailBrowsers{{44}}

\def\totalHTTPSFailAlt{{26}}

\def\totalHTTPSFailGPlay{{18}}

\def\httpsFailPercentage{{10\%}}

\def\HTTPSFailWarningBrowsers{{3}}

\def\HTTPSFailWarningPlay{{1}}

\def\totalHTTPSFailGPlayDownloadsM{{11}}

\def\certFailClaimSecureGPlay{{2}}

\def\quicBrowsers{{8}}

\def\pinningBrowsers{{6}}

\def\jsPermissionsSuccessful{{216}}

\def\jsPermissionsSuccessfulPercentage{{51\%}}

\def\jsPermissionsAPISupported{{4}}

\def\jsBatteryOnly{{216}}

\def\jsBatteryOnlyPercentage{{51\%}}

\def\jsBatteryAndWebAndXReq{{204}}

\def\jsBatteryUnknown{{8}}

\def\blockAndMiss{{195}}

\def\blockAndPII{{63}}

\def\blockAndShareHistory{{38}}

\def\blockAndCertFail{{20}}

\def\oneLib{{63\%}}

\def\totalHistoryLeakBrowsers{{81}}

\def\historyLeakPercentage{{19\%}}

\def\historyLeakersFeatures{{60}}

\def\historyLeakFeaturesPercentage{{14\%}}

\def\historyLeakersWithPII{{13}}

\def\historyLeakPIIPercentage{{3\%}}

\def\historyLeakersNonFeatures{{37}}

\usepackage{tikz}

\definecolor{lightgray}{rgb}{0.95, 0.95, 0.95}
\definecolor{darkgray}{rgb}{0.4, 0.4, 0.4}
\definecolor{editorGray}{rgb}{0.95, 0.95, 0.95}
\definecolor{editorOcher}{rgb}{1, 0.5, 0} 
\definecolor{editorGreen}{rgb}{0, 0.5, 0} 
\definecolor{orange}{rgb}{1,0.45,0.13}		
\definecolor{olive}{rgb}{0.17,0.59,0.20}
\definecolor{brown}{rgb}{0.69,0.31,0.31}
\definecolor{purple}{rgb}{0.38,0.18,0.81}
\definecolor{lightblue}{rgb}{0.1,0.57,0.7}
\definecolor{lightred}{rgb}{1,0.4,0.5}
\usepackage{upquote}
\usepackage{listings}
\lstdefinelanguage{CSS}{
  keywords={color,background-image:,margin,padding,font,weight,display,position,top,left,right,bottom,list,style,border,size,white,space,min,width, transition:, transform:, transition-property, transition-duration, transition-timing-function},	
  sensitive=true,
  morecomment=[l]{//},
  morecomment=[s]{/*}{*/},
  morestring=[b]',
  morestring=[b]",
  alsoletter={:},
  alsodigit={-}
}

\lstdefinelanguage{JavaScript}{
  morekeywords={typeof, new, true, false, catch, function, return, null, catch, switch, var, if, in, while, do, else, case, break},
  morecomment=[s]{/*}{*/},
  morecomment=[l]//,
  morestring=[b]",
  morestring=[b]'
}

\lstdefinelanguage{HTML5}{
  language=html,
  sensitive=true,	
  alsoletter={<>=-},	
  morecomment=[s]{<!-}{-->},
  tag=[s],
  otherkeywords={
  >,
	<!DOCTYPE,
  </html, <html, <head, <title, </title, <style, </style, <link, </head, <meta, />,
	</body, <body,
	</div, <div, </div>, 
	</p, <p, </p>,
	</script, <script,
  <canvas, /canvas>, <svg, <rect, <animateTransform, </rect>, </svg>, <video, <source, <iframe, </iframe>, </video>, <image, </image>, <header, </header, <article, </article
  },
  ndkeywords={
  =,
  charset=, src=, id=, width=, height=, style=, type=, rel=, href=,
  fill=, attributeName=, begin=, dur=, from=, to=, poster=, controls=, x=, y=, repeatCount=, xlink:href=,
  margin:, padding:, background-image:, border:, top:, left:, position:, width:, height:, margin-top:, margin-bottom:, font-size:, line-height:,
  transform:, -moz-transform:, -webkit-transform:,
  animation:, -webkit-animation:,
  transition:,  transition-duration:, transition-property:, transition-timing-function:,
  }
}

\lstdefinestyle{htmlcssjs} {%
  basicstyle={\footnotesize\ttfamily},   
  frame=b,
  xleftmargin={0.75cm},
  numbers=left,
  stepnumber=1,
  firstnumber=1,
  numberfirstline=true,	
  identifierstyle=\color{black},
  keywordstyle=\color{blue}\bfseries,
  ndkeywordstyle=\color{editorGreen}\bfseries,
  stringstyle=\color{editorOcher}\ttfamily,
  commentstyle=\color{brown}\ttfamily,
  language=HTML5,
  alsolanguage=JavaScript,
  alsodigit={.:;},	
  tabsize=2,
  showtabs=false,
  showspaces=false,
  showstringspaces=false,
  extendedchars=true,
  breaklines=true,
  literate=%
  {�}{{\"O}}1
  {�}{{\"A}}1
  {�}{{\"U}}1
  {�}{{\ss}}1
  {�}{{\"u}}1
  {�}{{\"a}}1
  {�}{{\"o}}1
}
\lstdefinestyle{py} {%
language=python,
literate=%
*{0}{{{\color{lightred}0}}}1
{1}{{{\color{lightred}1}}}1
{2}{{{\color{lightred}2}}}1
{3}{{{\color{lightred}3}}}1
{4}{{{\color{lightred}4}}}1
{5}{{{\color{lightred}5}}}1
{6}{{{\color{lightred}6}}}1
{7}{{{\color{lightred}7}}}1
{8}{{{\color{lightred}8}}}1
{9}{{{\color{lightred}9}}}1,
basicstyle=\footnotesize\ttfamily, 
numbers=left,               
numbersep=5pt,              
tabsize=4,                  
extendedchars=true,         %
breaklines=true,            
keywordstyle=\color{blue}\bfseries,
frame=b,
commentstyle=\color{brown}\itshape,
stringstyle=\color{editorOcher}\ttfamily, 
showspaces=false,           
showtabs=false,             
xleftmargin=17pt,
framexleftmargin=17pt,
framexrightmargin=5pt,
framexbottommargin=4pt,
showstringspaces=false,      
}%
\begin{document}

\title{Not Your Average App: A Large-scale Privacy Analysis of Android Browsers}
\author{Amogh Pradeep}
\affiliation{%
  \institution{Northeastern University}
  \city{Boston}
  \country{USA}
}
\author{Alvaro Feal}
\affiliation{%
  \institution{IMDEA Networks Institute / Universidad Carlos III de Madrid}
  \city{Madrid}
  \country{Spain}
}
\author{Julien Gamba}
\affiliation{%
  \institution{IMDEA Networks Institute / Universidad Carlos III de Madrid}
   \city{Madrid}
   \country{Spain}
}
\author{Ashwin Rao}
\affiliation{%
  \institution{University of Helsinki}
  \city{Helsinki}
  \country{Finland}
}
\author{Martina Lindorfer}
\affiliation{%
  \institution{TU Wien}
  \city{Vienna}
  \country{Austria}
}
\author{Narseo Vallina-Rodriguez}
\affiliation{%
  \institution{IMDEA Networks Institute / AppCensus Inc.}
  \city{Madrid}
  \country{Spain}
}
\author{David Choffnes}
\affiliation{%
  \institution{Northeastern University}
  \city{Boston}
  \country{USA}
}

\renewcommand{\shortauthors}{Pradeep et al.}

\begin{abstract}

The privacy-related behavior of mobile browsers has remained widely 
unexplored by the research community. 
In fact, as opposed to regular Android apps, mobile browsers may present 
contradicting privacy behaviors. On the one hand, they 
can have access to (and can expose) a unique combination of 
sensitive user data, from users' browsing history to permission-protected
\textit{personally identifiable information} (PII) such as unique identifiers
and geolocation. On the other hand, 
they are in a unique position to protect users'
privacy by limiting data sharing with other parties by 
implementing ad-blocking features. 
%

In this paper, we perform a comparative and empirical analysis on 
how hundreds of Android web browsers protect 
or expose user data during browsing sessions.
To this end, we collect the largest dataset of Android browsers to date, from
the Google Play Store and four Chinese app stores.
Then, we develop a novel analysis pipeline that combines static
and dynamic analysis methods to find a wide range of privacy-enhancing 
(e.g., ad-blocking) and privacy-harming
behaviors (e.g.,  sending
browsing histories to third parties, not validating TLS certificates, and
exposing PII---including non-resettable identifiers---to third parties)
across browsers.
We find that various popular apps on both Google Play and Chinese stores have
these privacy-harming behaviors, including apps that claim to be
privacy-enhancing in their descriptions.
Overall, our study not only provides new insights into important yet overlooked
considerations  for browsers' adoption and transparency, but 
also that automatic app analysis systems (\eg{} sandboxes) need  
context-specific analysis to reveal such privacy behaviors. 
\end{abstract}


\maketitle

\sloppy

\section{Introduction}
\label{sec:introduction}

Mobile browsers (\ie{} apps that allow users to visit websites) are complex, powerful, and 
poorly understood software systems that account for 55\% of global website
visits~\cite{mobileTraffic2}. 
Their critical role as one of the primary gateways to the web, and the rich set
of features that they support,
make mobile browsers a particularly interesting platform to study 
from a privacy perspective. 
On the one hand, mobile browsers can \emph{enhance} user privacy 
in unique ways by implementing features such as blocking web trackers
and advertisers,
enforcing secure network protocols wherever possible, and minimizing personal data 
exposure~\cite{nithyanand2016adblocking,privacyBrowsers,
  blockBrowsers, li2007spyshield, yu2016tracking}.
However, they can also inflict privacy harms by harvesting and exposing 
permission-protected information such as unique identifiers or user
geolocation to third parties (as is commonly found in non-browser apps), or 
indirectly by making them available to website scripts via JavaScript APIs.
Further, they may expose browser-specific sensitive data such as 
users' browsing history or credentials
to third-parties due to poor design choices or 
a need to generate revenue, potentially at the expense of
user privacy~\cite{razaghpanah2018apps, reyes2018won, englehardt2016online, yu2016tracking}. 

Despite their potential for harm, the research community has largely overlooked
the privacy threats inherent to mobile browsers. 
Early studies focused on a small set of browsers~\cite{198599,wu2014analyzing}
and identified isolated cases of ``privacy protecting''
browsers deceiving their userbase and abusing their access to
personal and browsing data for tracking purposes~\cite{braveCrypto, duckduckgoIcons}.
In this paper, we augment the state-of-the-art by 
conducting the first large-scale, systematic, 
and multidimensional analysis 
of the privacy behavior of \totalBrowsers{} Android browsers
available in public app markets (including the Google Play Store
and four Chinese markets) and others pre-loaded by
certain phone vendors.
Specifically, we study and characterize:
(1) how mobile browsers help or harm users' privacy during the course of
web browsing sessions;
(2) what additional permission-protected personal data mobile browsers collect
and share with other parties, and the implications of such data collection;
and (3) how the combination of these behaviors impacts the 
overall privacy disposition of the mobile browsers in our dataset.

While there is a significant amount of work on mobile app privacy space 
in general, the study of 
mobile \emph{browsers} poses unique methodological challenges that we
address in this work. 
First, neither the Android OS nor app markets provide a comprehensive 
way to determine whether an app is a mobile browser. 
To address this challenge, we combine code inspection and dynamic analysis
methods to identify those apps that match our definition of a mobile browser.
Second, both browsers and websites expose data to other parties while in use, 
and simple analysis of network traffic traces is insufficient to distinguish whether the 
data was exposed by a browser or a website. 
To address this, we develop a novel browser analysis pipeline that uses webpage 
replays and a \emph{baseline browser} to
attribute data collection to either the website or the browser itself.
We further use a combination of static and dynamic analysis techniques to identify which  
browser component (for instance, a third-party library) is responsible for data
collection and gather actual evidence of these behaviors.
Third, browsers can implement privacy-enhancing features to protect users from  
harm (\eg{} use of TLS encryption, limited access to JavaScript APIs
that retrieve sensitive information, 
blocking websites from contacting third parties), but there are limited standard 
benchmarks to test for them. 
To overcome these challenges, 
we build a new test suite that addresses this need, along with instrumentation and automation 
to capture and study browsers' behaviors automatically and at scale.

We use the above methods to analyze a dataset 
of \totalBrowsers{} browsers and observe the following:

\smallskip
\begin{packeditemize}

\item We find that \easylistBrowsersPercentage{} of browsers enhance privacy by
  blocking tracking scripts by default. Similarly, \jsBatteryOnlyPercentage{}
  of browsers block scripts that access protected JavaScript APIs.

\item We see that most browsers do not default to HTTP\textbf{S} (only \domainHttpsBrowsersPercentage{}
  of browsers do so) and that \httpsFailPercentage{} of browsers do not properly
  validate TLS certificates, making them vulnerable to person-in-the-middle attacks.
  
\item We find that \oneLib{} of browsers contain at least one
  third-party library related to advertisement and tracking,
  and that these
  libraries are often responsible for browsers requesting ``dangerous'' permissions.
  Furthermore, our run-time behavior analysis shows that \leakPIIPercentage{} browsers share
  PII with other parties over the Internet, and that \historyLeakPercentage{} browsers
  do the same for browsing history. While \historyLeakFeaturesPercentage{}
  of these browsers share this information for a specific feature (such as web search
  APIs), we find that \historyLeakPIIPercentage{} send it alongside
  personal data (thus harming user privacy).
  We also show that browsers often share both resettable and non-resettable
  identifiers with third-party servers across the Internet, including
  four instances of ID bridging, a practice that completely defeats the use of
  resettable identifiers.
  
\item We conduct a multidimensional analysis of individual browsers to understand 
    their overall privacy disposition. We find that few browsers uniformly improve privacy 
    (\eg{} FOSS Browser), while many exhibit multiple harms (\eg{} Yandex, Baidu, Opera \etc{}).
    We also find mixed behaviors: of the 
    \trackerBlockers{} (\percTrackerBlockers{}) browsers that block tracking content, we see that
    70\% also allow tracking requests, 
    23\% expose PII, \nv{Talk about android id bridging}
    14\% share browsing history, and
    7\% fail to validate certificates.  
\end{packeditemize}

\smallskip\noindent
Our study has important implications for (1) end
users who adopt a non-default web browser on Android, and for (2) automatic app analysis processes.
For the former, our results can help guide their decisions about which browser to install
from a privacy standpoint. 
In fact, understanding the privacy risks of the mobile browser
ecosystem is critical in the EU 
as Google Chrome is no longer the default browser on Android 
and EU citizens can choose any browser when configuring their 
device~\cite{chromePreinstalled, choiceScreen}.
For the latter, we show that existing sandboxes may be 
ineffective for studying the privacy risks of mobile browsers, and they 
can benefit from our novel methodology to analyze browser apps and 
gain actual visibility into their behavior.
We reported our findings to Google,
which is currently investigating potential corresponding breaches of their 
policies. We also responsibly disclosed observed security vulnerabilities to developers. 
To support reproducibility and foster further 
research in this area, we make our code and data publicly available at \url{https://github.com/NEU-SNS/mobile-browser}.

\section{Threat \& Protection Models}
\label{sec:background}

This section describes our privacy threat and protection models.
Our models are motivated by the fact that mobile browsers 
occupy a privileged position in terms of the type of data they can access:
(1) all Android devices are expected to have at least one mobile 
browser~\cite{AndroidCDDSoftware,gamba202preinstalled};
(2) like other apps, mobile browsers can access permission-protected 
device sensors, system resources and unique identifiers; and
(3) being endpoints for web traffic, they have access to all data from web
browsing \eg{} page content or browsing history.
Note that our study focuses on browser behavior in the default mode, as opposed to 
``safe'', ``private'', or ``incognito'' modes.
Previous work has shown that these modes have their own
security and privacy issues~\cite{aggarwal2010analysis}, and that they do not
prevent browsers from collecting user data~\cite{incognito, chrome-lawsuit}.

\subsection{Privacy Threat Model}
\label{sec:privacy_threat_model}
We assume that a benign browser should render webpages,
follow best practices for connection security, and adhere to data
minimization principles when it comes to exposing user data. We 
acknowledge that there may not be such a browser,
but we assume one for the sake of comparing against our threat model. 
We consider that a browser is ``privacy harmful'' if it deviates from this benign
browser model by exhibiting
any of the following behaviors:

\para{Data dissemination not required for page rendering.}
Browsers may collect and share sensitive data (\eg{} location, unique identifiers),
with first parties (the browser developer) or third parties
(\eg{} data brokers, advertisers, analytics companies).
Such sharing might be required to implement site features
(\eg{} geofencing or localized search results); or used for secondary purposes
(\eg{} for monetization through ads and tracking). \\
In this paper, sensitive data includes \emph{personally identifiable information} (PII) 
and users' browsing history. 
Specifically, we consider the following data types to be PII: 
IMEI, Advertising ID (AdID), Android ID, MAC 
address, and geolocation.
Note that the IMEI, Android ID, and MAC addresses are non-resettable IDs, 
while the AdID can be reset by users. We also consider the list of installed apps to be sensitive, as it can be used to profile users~\cite{seneviratne2014predicting}.
For browsing history leaks, we consider cases where over half of the websites visited are 
transmitted to another party over the Internet. 
While browsers may share the visited websites as part of their functionality
(\eg{} URL safety checks~\cite{google-safe-browsing, opera-sitecheck}), this data can also be abused to build unique and stable identifiers for
users, even across devices~\cite{bird2020history}.

\para{Website manipulation.}
Browsers may modify (\ie{} replace or add) content in a way that can harm user privacy.
Examples of this behavior are replacing referral links~\cite{braveCrypto} or
injecting content and web elements associated with advertising companies
that might collect user data
(\eg{} advertisements~\cite{thomas2015ad, arshad2016identifying}).

\para{Poor connection security.}
A failure to validate TLS certificates exposes users to 
person-in-the-middle attacks~\cite{razaghpanah2017studying}.
This can allow any arbitrary actor in the network to intercept all
traffic, thus damaging the user's privacy (as the data can be observed)
and security (data can be dropped or modified by the attacker).

\subsection{Privacy Protection Model}
\label{sec:privacy_protection_model}

We consider in our privacy protection model the following 
privacy-enhancing behaviors that are not  
implemented in the idealized benign browser 
behavior described above:

\para{Blocking tracking domains.}
A browser that blocks connections to advertising and tracking services and scripts
can potentially prevent third parties from collecting extensive data on 
users~\cite{acar2014web, acar2013fpdetective, englehardt2016online, yang2020comparative}.
We consider a browser to implement such privacy-protecting behavior if 
we can find evidence of blocked connections to such services.   

\para{Limiting access to WebAPIs.}
Websites may request access to sensitive data and sensors 
via JavaScript APIs (\eg{} the geolocation API). 
A privacy-protecting browser would block such requests by default. This includes cases 
where access is blocked until the user gives explicit permission to allow access.

\para{Upgrading connection security.}
Many websites support both HTTP and HTTPS, with the 
latter being preferred for privacy and security. Privacy-protecting browsers  
should always upgrade to HTTPS when supported by the server.

\section{Mobile Browser Dataset}
\label{sec:dataset}

As most app stores do not consider a browser as a specific app category, 
we need to develop a sound method to automatically collect and
identify mobile browsers at scale. 
However, this task is not trivial. 
Many Android apps rely on WebViews to show online content
(\eg{} ads~\cite{ads_webview}) or provide in-app browsing functionality often part of social networks~\cite{krause-tiktok,zhangh21inapp}. Thus
we need a way to differentiate those from actual browsers. 
In this section, we present our definition of a functional 
browser and describe our methodology to
obtain a representative dataset of browsers from different sources.

\subsection{Definition and Installation Sources}
\label{sec:definition}

\para{What is a browser?}
We define a browser as an app capable of accessing arbitrary
websites using HTTP/S URLs. More specifically, these apps must declare this capability 
to the Android OS via \emph{Intent filters} that handle \emph{URL schemes}.
The Android OS uses this same Intent-based filtering to display available browser
options when users click a URL, supporting our definition.
Our definition rules out both web-based mobile apps that render app-specific web
pages and apps that render webpages through links within them (via WebViews),
as they do not allow users to enter \emph{arbitrary} URLs.

\para{Where are browsers installed from?}
Android allows users to choose a default browser, which can either come
pre-installed on the device, or
can be downloaded by the user from the Google Play store.
Android can
also be configured to install apps from sources outside of Google Play.
We compiled a list of such alternate stores and 
their corresponding popularity
metrics (where available) and found that Chinese app stores stood out due to
their wide adoption~\cite{appinchina}.
Thus, in this study, we include apps from these different sources: the Google Play
Store, four popular Chinese app stores, and pre-installed browsers
present in Android devices from multiple vendors.
We believe this combination of sources captures apps used by most Android users
worldwide.

\subsection{Data Collection and Filtering}

\para{Browser collection.}
Collecting browsers from the sources listed above is challenging because most
app stores lack a ``browser'' category.
Thus, we use multiple strategies to collect \emph{potential}
browsers and filter them post collection to remove non-browsers.
We collected apps from all sources in September 2021.
First, we curate a list of \wellKnownBrowsers{} well-known browsers from
multiple online sources including newspapers, blogs, and prior
work~\cite{luo2017hindsight}.
We download the latest versions of these browsers from Google Play (accessed
from the EU).
For two Chinese stores, 360~\cite{360store} and Tencent~\cite{tencentstore}, we
use the search term ``browser'' in Chinese and collect all apps found with it.
The other two Chinese stores, Anzhi~\cite{anzhi} and AppChina~\cite{appchina}
provide a dedicated ``browser'' category, thus we collect all apps listed here.
Lastly, we include potential browsers for a dataset of pre-installed apps
gathered by Gamba~\etal~\cite{gamba202preinstalled}.
For this set, we look for the string ``browser'' in the package name and app
name (in the Android Manifest) to identify potential browsers.

\para{Filtering non-browsers.}
Our data collection strategy might yield apps that are non-browsers, \eg{} the
file explorer ``Moto File Manager'' passes our
string-based filtering as its package name (\texttt{com.lenovo.FileBrowser2})
contains the word \textit{browser}.
To filter such cases, we need a test that reliably 
determines whether a potential
browser is consistent with our definition of a browser (\S\ref{sec:definition}).
To this end, we conduct a two-stage testing procedure on each app to identify browsers. 
First, we determine whether the app successfully installs on a test device
running Android 11. If so, we then automatically trigger an \emph{Intent} through
the Android Debugging Bridge (adb) tool to drive it to open a test URL.
It is possible for an app to handle this Intent but
fail to load webpages due to run-time errors, or bad Intent-handling logic;
we thus validate this approach using dynamic analysis techniques
in \S\ref{sec:dynamic_pipeline_analysis}.
After the above filtering, our final dataset contains
\totalBrowsers{} browsers.
To further contextualize our results, we group these browsers into three categories
based on their origin: \totalBrowsersGPlay{} browsers
published on \emph{Google Play}, \totalBrowsersAlternate{} 
browsers published in \emph{Chinese stores} and \totalBrowsersPre{} browsers
that are pre-installed. This last number is small because many pre-installed
apps cannot run on our test device due to native or vendor-specific
dependencies~\cite{gamba202preinstalled, blazquez2021trouble}.
If we fetch the same browser from two different origins (\eg{} Google Play and
a Chinese store), we include it in both groups for our analysis.

\subsection{Identifying Unique Browsers}
\para{Uniqueness across origins.}
We want to avoid over-reporting by including the same browser (collected
from different sources) several times in our analysis. Grouping two apps can have certain
pitfalls due to the lack of robust methods for author attribution on Android.
While Google Play ensures that package names are unique \emph{within the
store}, this is not the case for apps from a different origin.
Thus, a package name on Google Play might be used by a different app on another
store (\eg{} a Chinese store).
In fact, prior research showed that apps sharing package names could be repackaged and thus
may have no relation to the original developer~\cite{khanmohammadi2019empirical}.
Further, in the case of pre-installed browsers, any manufacturer can modify and
pre-install any open-source browser app with the default package name, \ie{}
\texttt{com.android.browser}~\cite{gamba202preinstalled}.

Our key observation is that every app must be 
signed with a certificate to be installed on an
Android device~\cite{androidsigning, permsOverview}.
Thus, we rely on certificate information to distinguish two apps with the same
package name (assuming that the same app would not be signed by two
different certificates). We note that identifiers within self-signed 
certificates (\eg{} the \texttt{Subject} field), may be forged~\cite{lindorfer2014andrubis}.
Therefore, we do not associate a certificate with a particular developer, but rather use
information relating to the (unforgeable) private key to identify unique developers.
Specifically, we use the package name and SHA-256 of the signing certificate
combination as a unique identifier for each browser app.
This allows us to identify browser apps with the same package name
that are potentially implemented by different entities, and thus might behave
differently. 
We account for different signatures for the same app by looking at app hashes
after stripping certificate information from them and deduplicating matches.
We argue that our technique is a reasonable heuristic (without ground truth) to
uniquely identify browsers across origins.

\para{Uniqueness of browsers.}
In our dataset, the majority (\onlyOneHash{}) of browsers with the same package name 
are signed by the same certificate. However, we also find browsers with the same package name 
being signed with \emph{different}
certificates, up to \maxCerts{} different ones
(\eg{} \texttt{com.baidu.browser.apps} found in different Chinese stores).
Anecdotally, we also find versions of Firefox signed with different
certificates in Chinese stores and the Google Play Store.
This shows that some apps might be developed by different
companies depending on the store where they are available. 


\begin{table}[!t]
  \centering
  \caption{Most popular browsers on Google Play
    (100M+ downloads), Anzhi (1M+ downloads), and AppChina (Top 5).}
    
  \setlength{\tabcolsep}{3pt}
  \rowcolors{2}{gray!25}{white}
  \resizebox{.8\linewidth}{!}
  	{%
	    \begin{tabular}{clrcc}
	    \rowcolor{white}
	    &  \textbf{Name} & \textbf{\# DL GPlay} & \textbf{\# DL Anzhi} & \textbf{Top5 AppChina} \\
	    
	    \midrule
              \includegraphics[width=15px]{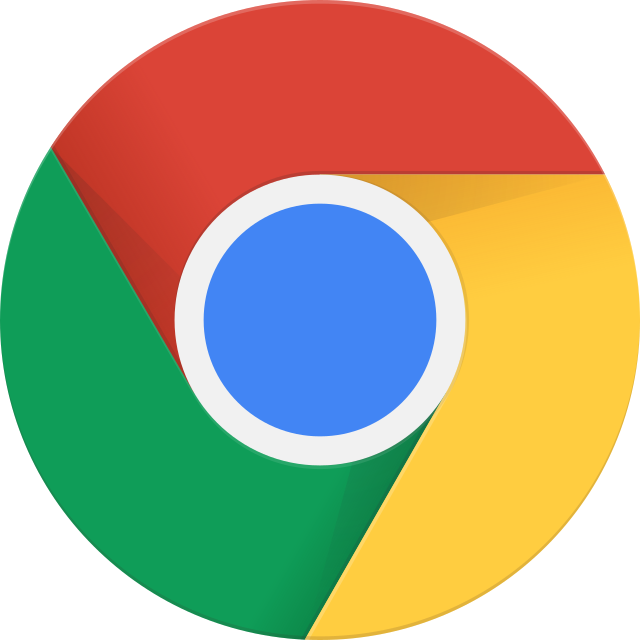} & Google Chrome & 10B+ & - & \no{} \\
              \includegraphics[width=15px]{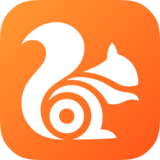} & UC Browser & 1B+  & 200M+ & \yes{} \\
	    \includegraphics[width=15px]{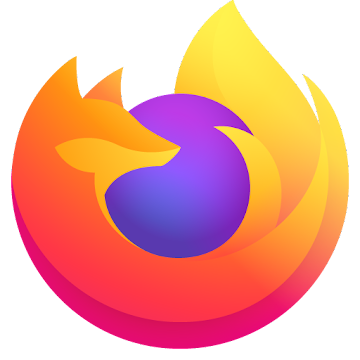} & Mozilla Firefox & 100M+ & 1M+ & \no{}   \\
	     \includegraphics[width=15px]{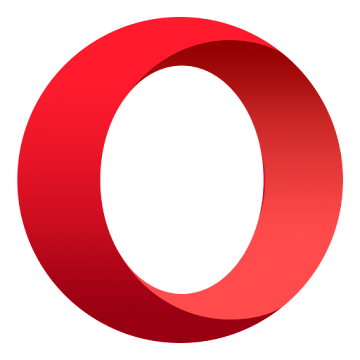} & Opera Browser & 100M+ & 7M+ & \no{}   \\
	     \includegraphics[width=15px]{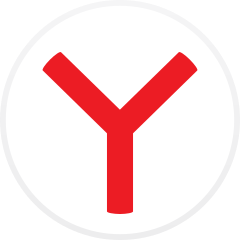} & Yandex Browser & 100M+ & -  & \no{}  \\
	     \includegraphics[width=15px]{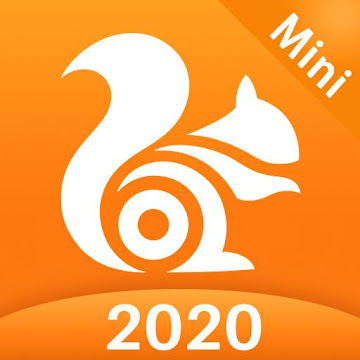} & UC-Mini & 100M+ & - & \no{}  \\
	     \includegraphics[width=15px]{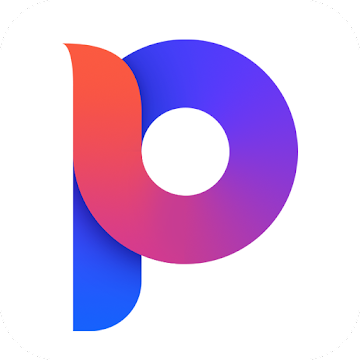} & Phoenix Browser & 100M+ & - & \no{}   \\
	    \bottomrule
	    
	    	    
	    \includegraphics[width=15px]{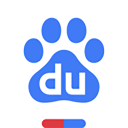} & Baidu & 5M+ & 100M+  & \yes{}  \\
	    \includegraphics[width=15px]{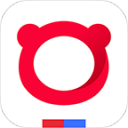} & Baidu Browser & - & 80M+ & \yes{} \\
	    \includegraphics[width=15px]{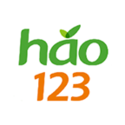} & Hao 123 & -  & 10M+ & \yes{} \\
	    \includegraphics[width=15px]{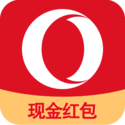} & Opera Extreme & -  & 4M+ & \no{} \\
	    \includegraphics[width=15px]{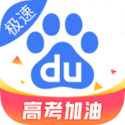} & Baidu Express & -  & 1M+ & \no{} \\
	     \bottomrule
	     
	     \includegraphics[width=15px]{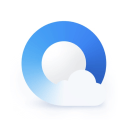} & QQ Browser & - & - & \yes{} \\
	    \end{tabular}
    }
  \label{table:top_10}
\end{table}

\subsection{Characterization of Browsers}

\para{Dataset statistics.}
Our final dataset contains \totalBrowsers{} browsers
with target API levels ranging from 5 (Android 2.0) to 30 (Android 11)
with a median of level 24 (Android 7.0).
While some of these target API levels might seem outdated, we note that
for every unique browser we fetch the latest available version.
This suggests that pre-loaded 
browsers and those distributed through app stores are not kept up to date,
and that they might
not benefit from the latest privacy-enhancing
techniques and security standards.

The size of our dataset (\totalBrowsers{} browsers) 
indicates that there is a diverse and vibrant market for mobile browsers,
providing a wide range of features. 
To better understand the advertised features and compare them with
actual browser behavior observed during dynamic analysis,
we extract and use as a proxy the most frequently used words in their app market 
descriptions (\percGPlayBrowsers{}
of our dataset). 
We find that \performanceBrowser{} of the browsers advertise themselves as
\emph{high performance} browsers---using words such as ``Fast'', ``Quick'' in their
description, \privacyBrowser{} use words related to \emph{privacy and security},
\priceBrowser{} advertise their \emph{low cost}---with words such as
``Free'', and \kidBrowser{} claim to be \emph{child-friendly} browsers. 
Regarding their market share, we find that \moreOneMBrowsers{} browsers
have over 1M downloads on Google Play, and 6 browsers have over 100M 
downloads. More broadly, we find 4 browsers with fewer than 10k downloads, 
65 apps have between 10k and 500k downloads, and 85 apps have over 500k downloads.
While we do not have a download figure for all of Chinese stores, both
Anzi and AppChina have a browsers category in which the available apps are
ordered by popularity. The differences between these stores and Play is notable
as the five most popular browsers on AppChina are the QQ Browser,
UC Browser and three products from Baidu: Baidu Search, Baidu Browser and
Hao 123. In the case of Anzhi, browsers with over 1M downloads include: UC
Browser, Baidu Search, Hao 123, Opera Browser, Opera Extreme, Baidu Express,
and Mozilla Firefox.
We list the most popular browsers published on Google Play, Anzhi and AppChina
in Table~\ref{table:top_10}.
These significant differences across stores, combined with the large numbers of estimated 
users (\eg{} QQ Browser and UC Browser accounting for more than a half billion monthly
active users~\cite{infNewsQQ}),
highlight the importance of including Chinese stores in our analysis.

\subsection{Browser Engine Attribution}
Web browsers are complex software that requires substantial engineering effort 
to develop from scratch.
While we cannot know for certain why there are so many browsers in the
Android ecosystem (\totalBrowsers{} according to our count),
one hypothesis is that many of them are built atop existing open-source 
browser engines (\eg{} Chromium, Gecko) or are built as a WebView
wrapper.\footnote{WebView objects allow developers to display web 
content on an activity layout. However, it lacks 
some of the features of fully-developed browsers like Chromium.} 
To investigate whether this hypothesis holds, we conduct a multi-facet analysis to infer whether a browser uses an underlying browser engine or is implemented using a WebView component provided by the Android OS~\cite{steiner2018web,beerbridge}.

\para{Class implementations.}
We search for standard classes for implementations of WebViews, Chrome,
Firefox, and ChromeCustomTabs. We note that, as any static analysis technique,
this might lead to false positives because of legacy and dead code.
\emph{WebView}-based browsers need to implement the
\texttt{android.webkit.WebViewClient} class to function, so we search for these
implementations and find that \webViewImplementers{} (\percWebViewImplementers)
browsers use them.
Similarly, we  find that \chromeBrowsers{} (\percChromeBrowsers) implement
Chrome related classes
(\texttt{{org.chromium.chrome.browser.ChromeTabbedActivity}}) and that
\firefoxBrowsers{} (\percFirefoxBrowsers) implement Gecko/Firefox related
classes (\texttt{org.mozilla.geckoview.GeckoSession}).
We also look for ChromeCustomTabs implementations
(\texttt{androidx.browser.customtabs.CustomTabsIntent}), which could in theory
be used to implement a browser~\cite{steiner2018web, beerbridge}, but found no
cases of it. We are unable to attribute the remaining
\engineUnknownBrowsers{} (\percEngineUnknownBrowsers) to any of these engines.

\para{X-Requested-With header.}
We complement the results from our static analysis with looking at the
headers sent during run-time requests made to test websites.
One of our key observations is that \emph{WebView}s set a
\emph{X-Requested-With} header to the package name of the
app while loading pages~\cite{webViewXRequestedWith}.
We see that
\webViewImplementerAndXRequested{} (\percWebViewImplementerAndXRequested) browsers
send this header for all test URL requests; \ie they use \emph{WebView}s to
render webpages users request.
All but \xRequestedNotWebView{} of the \emph{X-Requested-With} senders 
are also identified as \emph{WebView} implementers, providing a high level of 
cross-validation.
Manual inspection revealed that code obfuscation (preventing 
us from finding \texttt{android.webkit.WebViewClient}) is the likely 
reason for inconsistencies.

\para{User-Agent strings.}
We look at the \emph{User-Agent} strings advertised by browsers and match
them to possible underlying engines.
However, we can only attribute \userAgentWebViewBrowsers\ to \emph{WebView}s
with this technique.
Note that this technique has its own limitations, as \emph{User-Agent} strings
can easily be spoofed by the developers.

\para{Code similarity.}
As observed by related work, browser engines are likely written in native code and included as a shared library for performance reasons~\cite{wu2014analyzing}. Thus, as a last step, we tried to rely on the code similarity between every pair of browsers in our dataset to understand if they share the same browser implementation. Specifically, we compared the native libraries that are included in the apps using BinDiff~\cite{dullien2005graph,bindiff}, but did not
find conclusive results.

\section{Methodology}
\label{sec:methodology}

The goals of our analysis are: identifying privacy-harmful
and -enhancing behaviors in mobile browsers as outlined in \S\ref{sec:background}, 
determining the root causes for the behaviors we observe, and determining 
the impact of such behaviors on users' privacy.   
To meet these goals,
we rely on the complementary strengths of both static code 
analysis and dynamic analysis techniques. 

We note that static and dynamic analysis techniques have limitations 
when used on their own. 
Static analysis can cover many different execution paths without running 
apps, but it may produce false positives
resulting from legacy or dead code, and produce false negatives due to 
apps leveraging code obfuscation or loading code dynamically. 
Dynamic analysis reveals the impact of real code 
execution paths at run time, but has limited coverage,
can be defeated by anti-testing methods and provides a
lower-bound of all browser behavior due to its inability to trigger all
code paths with existing fuzzing methods~\cite{choudhary2015automated}.
The combination of both methods gives us better visibility for 
understanding browser functionality than using any one technique independently.
In addition to these techniques, when we identify potentially privacy-harming
behavior with dynamic analysis, we manually analyze the code to 
better understand and validate the logic triggering it.

\subsection{Static Analysis}
\label{sec:static-analysis}

Static code analysis allows us to understand the browsers' potential for 
privacy-invasive behavior.
Namely, this analysis reveals the types of data that an app can access
(\ie{} requested permissions), the types of third parties that are integrated in
browsers (\ie{} SDKs), and whether these SDKs are piggybacking on the
permissions requested by the browser to access sensitive data for secondary purposes.

\para{Permission analysis.}
The Android Open Source Project (AOSP) implements a permission model to restrict
access to some of its features and sensitive resources 
(\eg{} the user's location, or SMS messages), to protect
user privacy and security~\cite{permsOverview}.
However, these permissions are shared among the host app embedded third-party
SDKs.
These SDKs might access personal data for secondary purposes.
This might also lead to apps requesting
more permissions than strictly necessary~\cite{felt2011android, au2012pscout},
as SDKs often need access to
data that is not necessary for the functioning of the app or to SDKs
leveraging the set of permissions of the host app to access
sensitive resources (potentially without user consent). 

Thus, to assess the privacy risks of mobile browsers, we 
parse each browser's manifest to extract the requested permissions.
This approach provides an upper bound of permission protected data that 
can be shared with other parties. Then,  
we map API calls (\ie{} functions called within the browser's code) to the
AOSP permissions that protect them in order to gain a more fine-grained
perspective. Note that Google does not include such a mapping
in their documentation. Therefore, we leverage three complementary sources:
(1) the mappings implemented in Android Studio (Android's official IDE),
which contains scripts
to warn developers if they use an API without requesting the associated
permission~\cite{androidStudioAnnotations}; (2) mappings
extracted by parsing the AOSP source code to extract the methods that use the
\texttt{@RequiresPermission} annotation, part of the AndroidX and Android
support libraries~\cite{requiresPermAndroidX, requiresPermAndroid}; and (3)
the mappings generated by prior work
(Axplorer~\cite{backes2016demystifying}) for older Android versions (Android 5 and below). 
We acknowledge that, while we update the
mappings released by prior work, our mappings might still be
incomplete (something that we cannot measure due to the lack of ground truth).

\para{Third-party libraries.}
Android apps often rely on third-party libraries
(or SDKs) to integrate functionality 
offered by third-party entities, \eg{} for A/B testing, analytics, or
advertisements~\cite{razaghpanah2018apps}.
We use LibRadar~\cite{ma2016libradar} to identify third-party libraries in browser code
but, as its library fingerprints and library-to-company (LTC) mappings  
are outdated~\cite{feal2021don}, we augment LibRadar's LTC mappings
with data from Exodus~\cite{exodus} and other information gathered through
online resources and open-source intelligence.
We further use domain knowledge to identify the purpose of the library as well
as the company behind the service, and use it to
characterize the kind of third-party SDKs that browsers are using.

To further discern whether permission-protected APIs
are requested by first-party or third-party code, 
we use Androguard~\cite{androguard}.
Specifically, we use Androguard to extract all
of the permission-protected API calls in the code, extract the package of the class
invoking it, and then use LibRadar to label
whether this class belongs to the browsers' code or to a third-party
SDK. 
We note that some of the SDK's code might never be used, and thus we filter out
all third-parties for which we do not find a cross-reference in the
browser's code to reduce over-reporting.

\begin{figure}[t!]
\centering
\includegraphics[width=\linewidth]{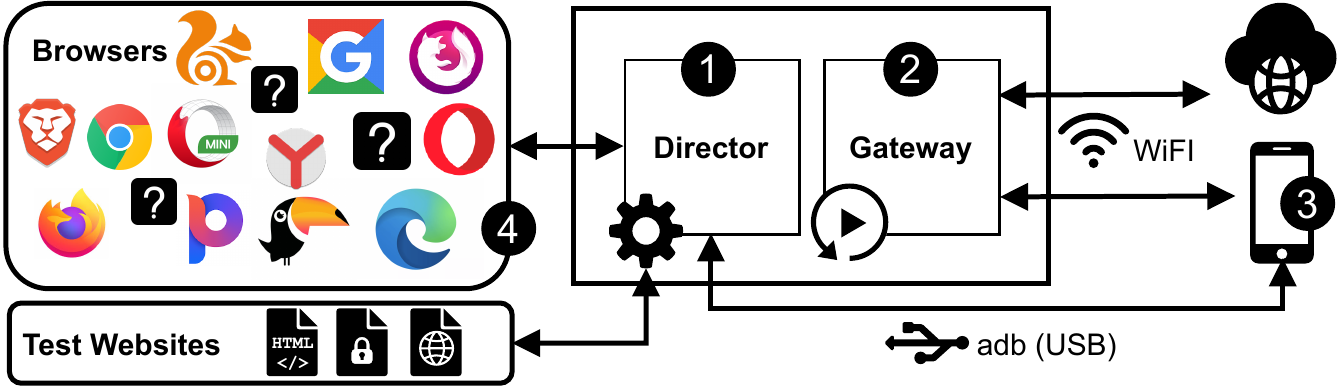}
\caption{Overview of our dynamic analysis pipeline.}
\label{figure:dynamic_analysis}
\end{figure}

\subsection{Dynamic Analysis}
\label{sec:dynamic_pipeline_analysis}

Dynamic analysis gathers evidence of browser behavior by executing 
browsers on an instrumented device and observing their behavior. Specifically, we monitor:
sensitive data exposure, network traffic security practices, and web content
manipulations as browsers fetch and load websites.

\subsubsection{Pipeline}

To perform dynamic analysis at scale, we develop an automated 
black-box testing instrumentation
and logging testbed, which consists of several components as shown in
Figure~\ref{figure:dynamic_analysis}.
Table~\ref{table:data_collected} (in Appendix) summarizes the data collected by each
component.

\para{Director.}
The \director{} (Fig.~\ref{figure:dynamic_analysis}~\dcircle{1}) 
drives and orchestrates the analysis by installing the browsers from our dataset 
(Fig.~\ref{figure:dynamic_analysis} \dcircle{4}) via \texttt{adb} 
on a physical mobile device (Fig.~\ref{figure:dynamic_analysis}~\dcircle{3}), 
instructing the device to open webpages using an \texttt{ACTION\_VIEW} Intent~\cite{androidOpenURL},
and uninstalling browsers via \texttt{adb} when tests are complete. 
The \director{} also collects device logs and screenshots
during tests.
The \director{} consists of JavaScript ($\approx$500 LOC) and bash ($\approx$400 LOC).

\para{Traffic interception and injection with the gateway.}
The Gateway (Fig.~\ref{figure:dynamic_analysis}~\dcircle{2}) serves as a Wi-Fi
hotspot for mobile devices.
We route all mobile device traffic to an instance of
\texttt{mitmproxy}~\cite{mitmproxy} using \texttt{iptables} rules, and log all
traffic with \texttt{tcpdump}.
Using \texttt{mitmproxy}, the \gateway{} intercepts all mobile-device traffic.
In combination with interception, we inject into each webpage a \texttt{tripwire.js} (inspired by
Reis~\etal~\cite{reis2008detecting}), which is a piece of
JavaScript code ($\approx$100 LOC) that collects the Document Object Model (DOM) of content
delivered to the browser.
We develop a Python \texttt{mitmproxy} add-on to perform these actions and use
\texttt{tcpdump} to collect traffic that \texttt{mitmproxy} does not handle.

\para{Test device.}
To enable the \gateway{} to modify traffic, we edit an Android 11 factory image
(Build RQ3A.211001.001) and include our \texttt{mitmproxy} certificate as a
root certificate on this image.
In previous versions of Android (7 to 9), we could modify the root store (on the
system partition) with root access; starting with Android 10 this is no longer possible 
because the system partition cannot be modified after installation.
Thus, modifying the image before installation is the most convenient way to
include a new root certificate.
We run this modified factory image on a Pixel 3 device and use it as our test 
device (Fig.~\ref{figure:dynamic_analysis}~\dcircle{3}).

\para{Assessing pipeline compatibility.}
Before conducting a large-scale analysis, we determine whether 
every browser in our dataset can run automatically (\ie{} without human intervention) 
in our testing pipeline.
To do so, we induce each browser to open a simple webpage (using an
Intent) hosted locally, and check whether the
browser issues a request for it.
We find that \loadedOneAuto{} browsers successfully pass this test without any
interaction.
In addition to these, we find that \loadedOneManual{} browsers require minimal
human interaction (dismissing onboarding screens) to be compatible with
the pipeline.
While testing these  \loadedOneManual{} browsers, we manually dismiss the 
onboarding screens so the automated tests can run.
The remaining \noTest{} browsers fail to follow the Intent to load pages and are
thus incompatible with our test.
These cases are indeed browsers, but fail to load a simple webpage 
automatically due to unexpected crashes or requiring users to
login to use them.
Nevertheless, we still include these browsers in our experiments, and report
any privacy-related behavior we observe while they run 
(\eg{} some browsers send personal data to servers as soon as they are launched,
even if they do not load a webpage).
The experiment analysis logic consists of $\approx$4,000 lines of Python code.

\subsubsection{Test Inputs}

To observe any privacy-related browser behaviors, one must use browsers as 
they are intended: \ie visiting a series of real websites.
Our testbed uses Intents to automatically induce browsers to
fetch web content from the Internet, as if it were requested by a real user.
We note that in our experiments, we do not click links inside the webpage,
rather waiting for the page to fully load. 
For completeness, we rely on the following types of webpages
(Appendix~\ref{appendix:website_selection}
provides more details): (1) a
\textit{honeypage} that contains several types of popular tracking 
services that browsers might block, (2) a \textit{permissions page} that tries
to access different device sensor APIs, (3) a \textit{domain without a protocol specified}
to study how browsers handles HTTP(S) by default, (4) a \textit{HTTPS webpage}
to analyze TLS security, and (5) 13 \textit{popular webpages}.
Of the \loadedOneTotal{} browsers that adhere to Intents and open webpages, we
find that \loadedAllPages{} load all of the test pages.

\para{Handling dynamic web content.}
To build our cache of site content, we load all the selected webpages 
using a \emph{baseline browser}. 
\footnote{We note that any trusted browser can serve as a
  baseline, but we use the default Android WebView for our Android 11 test device.}
Even with cached content, some webpages include non-deterministic
and dynamic content (\eg{} a timestamp in the HTML source or
ads resulting from real-time bidding processes~\cite{yuan2013real,10.1145/3366423.3380113}) 
that leads to different requests over time and impedes comparing the
same websites across browsers at different times. 
To address this challenge, we heuristically chose to rebuild the cache periodically 
(once every 50 browsers we test, which corresponds to about once every 2-3 hours).
For each (non-baseline) browser test, our testbed prepares the mobile device by
factory resetting it, installs necessary components, and collects identifiers 
(AdID and Android ID).
Moreover, to ensure consistency in webpages delivered, the \gateway{} caches copies
of tested websites and replays captured content using \texttt{mitmproxy}.

\para{Detecting browser manipulations.}
Browsers can use their privileged position to 
actively modify web content in ways that protect
(\eg{} ad-blocking) or harm users privacy (\eg{} JavaScript injection for
tracking purposes). We design 
a novel methodology to detect such changes by considering a combination 
of network requests seen across the browsers in our dataset, and instrumenting 
webpages to observe changes to each page's DOM. However, 
a key challenge is discerning whether such modifications
are caused by the browser or by dynamic
content as previously discussed. 
Therefore, 
our method for detecting content modification by browsers focuses on the 
most common known case, which is changing content in the DOM. 
Any other modifications that do not manifest in this way
 (\eg{} at rendering time) are out of scope. For that, 
we compare a \emph{baseline} DOM
of a webpage (assumed not to be modified) to the one rendered in the browser
under test.  
Specifically, we generate a baseline DOM and instrument future webpage tests by injecting a 
JavaScript-based tripwire that detects and reports DOM changes compared to this baseline.  
To compare the browser DOM to our \emph{baseline} DOM, we look at two
particular HTML tags: \texttt{script} and \texttt{link} 
to find both added and missing elements.
For each website in our test, we manually identify legitimate elements that
vary across crawls and automatically remove them from the collected DOMs.

\subsubsection{Network Traffic Analysis}
We log all network traffic generated by browsers using \texttt{tcpdump},
and our testbed intercepts and injects/replays all HTTP/HTTPS traffic with our \gateway{} to
serve consistent traffic and collect DOMs as described previously. However, 
we filter out flows that are not suitable for our webpage replay system. 
We drop all UDP traffic that is not DNS (port 53), including QUIC traffic. 
We found \quicBrowsers{} browsers that use QUIC at least once and have no visibility into this 
traffic. Our approach also has no visibility into HTTPS traffic that uses any form of 
certificate pinning that \texttt{mitmproxy} cannot intercept. While pinning in browsers is a largely obsolete technique~\cite{pinning:imc22}, we still find the case of \pinningBrowsers{} browsers that implement some for of pinning \emph{and} that we cannot bypass.
Given these limitations, the information exposure we do find serves as a lower bound.

\section{Browser Functionality}
\label{sec:browser_functionality}

We now analyze whether browsers protect or expose sensitive data  
during browsing sessions, according to our privacy protection and threat models
in \S\ref{sec:background}.
Namely, we analyze whether browsers modify the content of a website that is served
to users. This can potentially improve privacy by 
blocking requests to services known to track and profile users (\eg{} analytics
services and ad networks) or to harm privacy (\eg{} when a 
browser adds tracking code to a website).
We further analyze whether browsers block access to protected JavaScript APIs
that can be used to gather sensitive data about the user and the device. 
Finally, we determine whether browsers help or harm privacy with respect 
to connection security/privacy.

\subsection{Content Modification}
\label{sec:content_modification}
\para{DOM-based blocking.} We use our DOM-based tripwires to investigate which
content is blocked by browsers and the potential privacy impact of this
behavior. 
We begin by identifying cases that are highly likely to correspond to 
tracking services, \ie{} the content matches an entry in at least one of 
two popular Adblock filters 
(\texttt{EasyList}~\cite{easylist-version} and
\texttt{EasyPrivacy}~\cite{easyprivacy-version}).
We find that \easylistBrowsers{} browsers in our dataset (\easylistBrowsersPercentage)
block content flagged by these blocklists, strongly indicating that they 
are enhancing privacy for users. 

Interestingly, most of the blocked content that we observe
(\nonEasylistBlockersPercentage{}) is not part of these lists.
After manual analysis, 
we identify common trends that we can directly link to privacy-related behavior.
Of the browsers blocking content not on these lists, we find that
\analyticsBlockersPercentage{} block analytics content, \adBlockersPercentage{}
block ad content, and \widgetBlockersPercentage{} block widgets (\eg{} an embedded Twitter feed).
We look at libraries included in browsers to see if a common library results in
this blocking, but found no evidence of such code.
While we do not know exactly why these browsers block this content,
we speculate that this could be an attempt to mitigate pervasive tracking from
webpages. This also highlights the diversity in browsers' blocking behavior,
indicating that different browsers use different anti-tracking and blocklist
implementations~\cite{zafar2021adblock,feal2021blocklist}.

\para{Allowed requests.}
In contrast to the above examples, browsers might impact privacy by changing and/or blocking 
\emph{network requests} during page loads (\ie{} without changing the DOM). 
While it may seem trivial at first to detect blocked requests by comparing with 
a baseline browser, the key challenge for detecting this behavior 
is the lack of ground truth as to why a request is not issued. For example, a 
request could be absent due to a browser blocking a tracking service, but it also 
could be due to dynamic webpage behavior not captured 
in our baseline (\eg{} non-deterministic advertisements 
resulting from real-time bidding). 
Given this issue, we focus our analysis instead on the privacy 
impact of network requests to destinations matching the Adblock filters
mentioned above. In this analysis, we focus simply on whether browsers seem to
be using these filters to block all corresponding requests.
If we see no requests at all to destinations on the Adblock filters, 
we assume the browser is using such filters to improve privacy.
While we acknowledge that not all browsers implement blocklist approaches, we rely
on these resources as ground truth to detect most browsers that are actively
protecting the privacy of users.

We attribute requests to webpage loads in our tests by first restricting our analysis 
to only those with \texttt{HTTP referer} headers in a chain rooted at the loaded webpages. 
Next, we search for these requests in the Adblock filters.
We find that \easylistButAllowed{} browsers (\easylistButAllowedPercentage)
permit at least one request that should be blocked based on these lists.
One possible explanation for this behavior is that browsers use 
different blocklists, unblock lists, or neither---again, consistent with 
prior work observing variations in blocking behavior~\cite{zafar2021adblock,feal2021blocklist}.
Of the remaining \easylistPossiblyBlocked{} browsers, we look at browsers that
load a majority of our test pages and make no requests on these lists and see
that \easylistDefinitelyBlocked{} browsers
(\easylistDefinitelyBlockedPercentage{}) fit this criteria, including popular
browsers like Firefox and Adblock Browser.

\para{Content injection.}
Our DOM-based tripwire detects four browsers that inject extra scripts 
into loaded webpages. For two of them, we cannot locate them and thus
we cannot asses their impact for privacy.
Namely, Zdllq loads a script from the local file system
that we could not locate even when de-compiling the app. The browser 
Yuyan injects a script downloaded 
from \emph{pr.shuk.cn} which we only observed when loading the 
\emph{office.com} and \emph{patreon.com} websites, and for which we could 
not identify the purpose of the script or the server it contacts. 
Orbitum injects an open-source
script called UseAllFive~\cite{allfive} that is related to UI functionality 
(and does not appear to be malicious). Both this browser and
Dingzai also inject highly obfuscated scripts 
that prevent us from assessing their privacy impact.
We argue that injecting remote scripts into a
webpage is generally problematic and a potential risk for users, 
particularly given the little transparency about their purpose.

\subsection{Blocking Access to Protected APIs}
\label{sec:blocking_apis}
We study how browsers enhance privacy by enforcing permissions 
when webpages invoke privacy-sensitive JavaScript APIs to access data. 
We do so by creating a test webpage that
contains JavaScript code to access user data through WebAPIs.
We test an extensive set of APIs including location, camera and
microphone.
We find that our test succeeds for \jsPermissionsSuccessful{} browsers
(\jsPermissionsSuccessfulPercentage); all of these browsers reveal
battery charging status, battery level and battery charging time without user
permission.
We also observe that \jsPermissionsAPISupported{} browsers support a
permissions API that prompts users for access to various sensors (magnetometer,
accelerometer, \etc).
The default behavior for the remaining browsers is to deny access to any device
sensors.

To understand why so many browsers block access to data from 
WebAPIs, we test the hypothesis that they could all simply be 
using default behavior built into a common implementation: Android \emph{WebViews}.
This is important as, by default, the \emph{WebView} prevents exposure of the
most sensitive data that has been shown to be useful for user tracking, allowing
access only to the device accelerometer, magnetometer and battery~\cite{webAPIsMozilla}.
Using the results from our browser engine attribution presented in \S\ref{sec:dataset}, we see that
of the \jsBatteryOnly{} browsers that protect access to these APIs,
\jsBatteryAndWebAndXReq{} implement \emph{WebView}s and send
\emph{X-Requested-With} headers.
The \jsPermissionsAPISupported{} that support the permissions API are most likely
to be Chrome-based, using this information (matches Chrome-based browsers).
Lastly, we manually check the remaining \jsBatteryUnknown{} and see that they
contain obfuscated code preventing our engine attribution.

\subsection{Connection Security}
\label{sec:connection_security}
We now describe how we conduct tests and capture TLS handshakes 
to understand whether browsers protect users' privacy and ensure
connection security by correctly using TLS-based protocols. 
We focus on certificate validation and default connection  
security.

\para{Default protocol preference.}
To determine whether browsers default to using the secure HTTPS protocol over
HTTP, our testbed induces a browser to visit
a domain without specifying the protocol to use
(see \S\ref{sec:dynamic_pipeline_analysis} and Appendix~\ref{appendix:website_selection}).
For privacy and security, a browser 
should always favor HTTPS~\cite{paracha2020deeper}.
However, we find that only \domainHttpsBrowsers{} 
(\domainHttpsBrowsersPercentage{}) of the browsers pick HTTPS by
default.
Of these browsers, just \domainHttpsOnPlay{} is available on the Google Play Store 
(FOSS Browser), the remaining \domainHttpsOnAlt{} browsers are from Chinese stores.
None of the most popular browsers on Google Play, Anzhi, or AppChina (see Table~\ref{table:top_10})
implement HTTPS by default.

\para{Certificate validation.}
Browsers that do not properly validate TLS certificates compromise user security
as they enable adversaries to
mount person-in-the-middle attacks as they would accept arbitrary certificates.
To detect whether such attacks are possible, we use
the methodology presented in
\S\ref{sec:dynamic_pipeline_analysis}, but skip installing the
\texttt{mitmproxy} root certificate on the mobile device.
As a result, our self-signed certificates used during TLS interception 
is not trusted by the system and the corresponding 
connections should be dropped.
We then visit an arbitrary HTTPS website with each browser on our test device; a correct
implementation of TLS validation should reject the certificate.
Nevertheless, we find that \totalHTTPSFailBrowsers{} (\httpsFailPercentage{})
browsers accept the invalid certificate and are thus vulnerable to arbitrary
TLS interception attacks. 
These browsers originate both from Chinese stores (\totalHTTPSFailAlt{}) and
Google Play (\totalHTTPSFailGPlay{}, downloaded 
\totalHTTPSFailGPlayDownloadsM{}M+ times combined).
Of the \totalHTTPSFailBrowsers{} browsers, \HTTPSFailWarningBrowsers{}
(\HTTPSFailWarningPlay{} from Google Play) display a
warning stating that there is a possible security problem. 
Nevertheless, while the warning is in place, the browsers finish loading the
webpage in the background.
Ironically, of the \totalHTTPSFailGPlay{} browsers that are
available on Google Play, \certFailClaimSecureGPlay{}
(InBrowser Incognito and InBrowser Beta both developed by ``Private Internet Access, Inc")
advertised in their description that they provided
users with ``secure'' and ``incognito'' features, yet they fail to validate
certificates.
We reached out to the 14 developers for which we could find contact information
and also opened a bug with Google to report this issue.

\para{Secure protocols.}
Finally, we look at how browsers affect users by relying on secure protocols
(namely, DoH, DoT and OCSP).
Our experiments revealed no use of DNS-over-HTTPS (DoH) or
DNS-over-TLS (DoT) by default, which would provide browsers with private/secure
DNS queries and responses.
Thus, all browsers' DNS requests are sent in plaintext (allowing ISPs 
and other network observers to see them). 
We also did not see any Online Certificate Status Protocol (OCSP) traffic, a
protocol that can be used to verify the revocation status of a X.509
certificate. Our results are consistent with prior work indicating that mobile browsers 
fare poorly when it comes to correct certificate validation~\cite{liu2015end}.

\section{Privacy Analysis}
\label{sec:pii_exposure}

As with any ordinary Android app, mobile browsers have access to PII, 
such as Android Advertisement IDs (AdID), persistent identifiers, or device
location, most of which are protected by Android permissions.
Further, some browsers might request permissions that are 
used for secondary purposes, \eg{} for embedded SDKs or to support 
JavaScript APIs that can in turn be used by code on arbitrary websites.
In addition, they have unique access to browsing history and browsing patterns.
In this section, we examine the permissions
requested by mobile browsers regardless of the purpose.
Then we identify PII in network traffic observed during dynamic
tests and analyze their privacy implications for users.
Finally, we analyze cases where browsers potentially harm user privacy 
by sharing browsing history.

\subsection{Permission Analysis}
\label{sec:perms}
To understand the type of access to user data that browsers have, we analyze the
permissions they request following the methodology discussed in \S~\ref{sec:static-analysis}.
We extract a total of \allRequestedPerms{} unique permissions requested by the
apps in our dataset.
This includes both permissions defined in the Android Open Source Project
(AOSP)~\cite{permsOverview} and custom permissions~\cite{defineCustomPerms}.
Table~\ref{tab:permissions_per_dataset} shows the number of requested 
permissions per browser origin.
We find that browsers that are pre-loaded or distributed through Chinese stores
tend to request more permissions than those listed on Google Play: the
median number of permission requests (AOSP and custom included) is
\medianAllPermsPI{} for pre-loaded browsers and \medianAllPermsAlt{} for
browsers from Chinese stores, compared to \medianAllPermsGP{}
for browsers from Google Play.
However, the maximum number of permissions requested by pre-installed browsers
is \emph{lower} than browsers from other sources: only \maxAllPermsPI{}
permissions, compared to \maxAllPermsGP{} for browsers available on Google
Play, and up to \maxAllPermsAlt{} for browsers in Chinese stores.
In total, we extract \nbCustomPerms{} custom permissions requested by the
apps in our dataset.
However, such permissions usually lack documentation~\cite{gamba202preinstalled},
and their purpose is difficult to automatically and reliably infer from the application itself.
We thus discard such permissions from the rest of our analysis.

\begin{table}[t]
  \centering
  \caption{Number of requested permissions per dataset.}%
  \resizebox{\columnwidth}{!}{%
    \begin{tabular}{lrrrrrr}
    \multirow{2}{*}{\textbf{Origin}} & \multicolumn{2}{c}{\textbf{AOSP perms.}}
                            & \multicolumn{2}{c}{\textbf{Custom perms.}}
                            & \multicolumn{2}{c}{\textbf{All perms.}} \\
      & Median & Max & Median & Max & Median & Max \\
    \midrule
    Chinese Stores & \medianAOSPPermsAlt{} & \maxAOSPPermsAlt{}
                & \medianCustomPermsAlt{} & \maxCustomPermsAlt{}
                & \medianAllPermsAlt{} & \maxAllPermsAlt{} \\
    Google Play Store & \medianAOSPPermsGP{} & \maxAOSPPermsGP{}
                & \medianCustomPermsGP{} & \maxCustomPermsGP{}
                & \medianAllPermsGP{} & \maxAllPermsGP{} \\
    Pre-installed  & \medianAOSPPermsPI{} & \maxAOSPPermsPI{}
                & \medianCustomPermsPI{} & \maxCustomPermsPI{}
                & \medianAllPermsPI{} & \maxAllPermsPI{} \\

    \midrule
    \textbf{All origins}& \textbf{\allOriginsMedianAOSPPerms{}} & 
\textbf{\allOriginsMaxAOSPPerms{}}
                & \textbf{\allOriginsMedianCustomPerms{}} & 
\textbf{\allOriginsMaxCustomPerms{}}
                & \textbf{\allOriginsMedianAllPerms{}} & 
\textbf{\allOriginsMaxAllPerms{}} \\
  \end{tabular}%
  }
\label{tab:permissions_per_dataset}
\end{table}

\begin{figure*}[t]
  \centering
  \includegraphics[width=\linewidth]{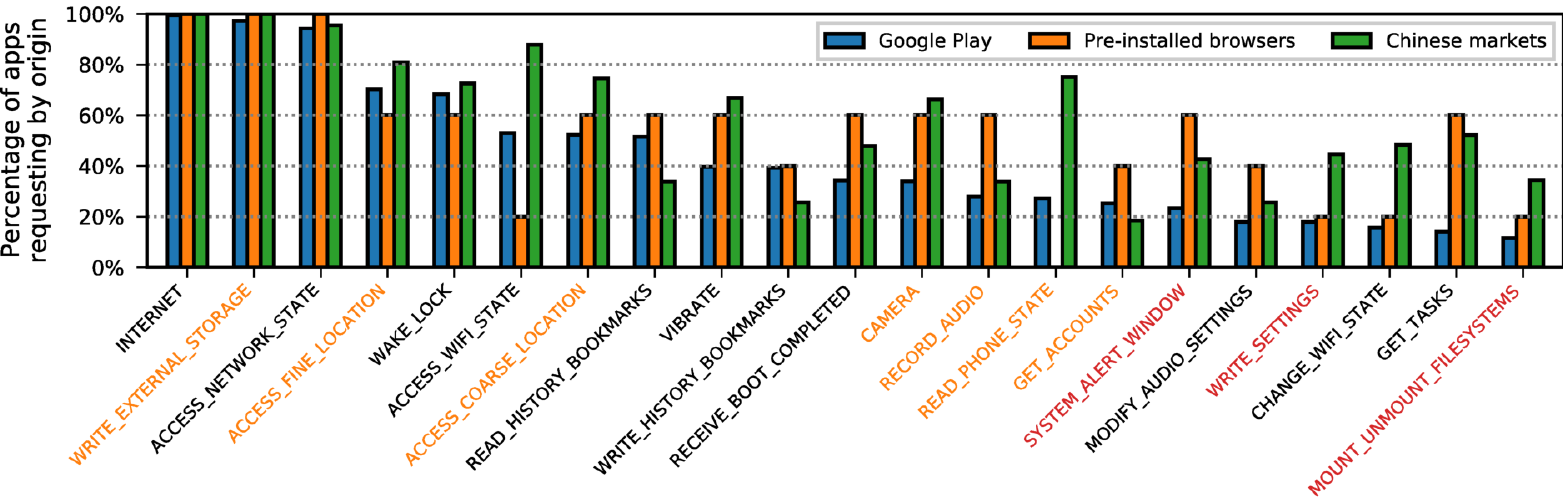}
  \caption{Frequently requested permissions: dangerous are in orange 
    and in red those not available to third-party developers.}
  \label{fig:most_requested_perms}
  \vskip-0.5em
\end{figure*}

Figure~\ref{fig:most_requested_perms} shows the number of apps
requesting AOSP permissions.
We show only the permissions requested by at least 20\% of the
apps in our global dataset to maintain readability.
The height of each bar indicates the percentage of apps, among the apps 
from the same origin, that request a given permission.
Permissions with a black label (\textbf{normal permissions})
are classified as ``low risk'' for users in
Android's AOSP documentation, they are granted at installation time
(\eg{} \texttt{INTERNET})~\cite{permsOverview}.
Similarly, permissions with an orange label (\textbf{dangerous permissions})
are those that must be explicitly granted at runtime as they can put the
user's privacy at risk (\eg{} \texttt{ACCESS\_FINE\_LOCATION}
gives access to the fine-grained geolocation of the device). 
Finally, permissions with a red label (\textbf{signature permissions}) are
automatically granted (with no user interactions) to an app if it is signed with
the same certificate as the app that declared the permission.
Only system apps (\ie{} apps located on one of the system partitions of
the device) can be granted such permissions~\cite{privilegedApps}, as they
allow an app to access very sensitive information (\eg{} the
\texttt{READ\_LOGS} permission allows an app to read the system logs) or to
perform low-level operations such as mounting and unmounting filesystems.
Some publicly available browsers request such permissions nevertheless.
One possible explanation is that these browsers can also come pre-installed on some devices
(\eg{} Google Chrome), and that developers do not remove these permission requests 
before submitting their apps to app markets.

Unsurprisingly, we find that the most frequently requested permission is
\texttt{INTERNET}, which is required for Internet access, followed by 
\texttt{WRITE\_EXTERNAL\_STORAGE}, for writing to the SD card,
and \texttt{ACCESS\_NETWORK\_STATE}, for checking the device's
connectivity.
These permissions are necessary for implementing basic features of a web browser.
Figure~\ref{fig:most_requested_perms} also includes now-deprecated
permissions that we keep in our analysis, as they still give
access to protected resources to apps
on devices running an old-enough Android version.
This includes \texttt{WRITE\_} and \texttt{READ\_} \texttt{HISTORY\_BOOKMARKS}
which, until Android 6.0, would allow apps to
``read (resp.\ modify) the history of all URLs that the Browser has visited,
and all of the Browser's bookmarks''~\cite{bookmarkPerms, bookmarkPermsDescs}.

Our results show that a large number of browsers also request access to
sensitive features, such as the location or recording audio. 
These permissions may have a legitimate use case, such as enabling
access to JavaScript APIs (\eg{} a weather website requesting the
user's location, or a videoconferencing site needing access to the microphone and
camera). In any case, the fact that the browser 
has access to this information can pose a privacy risk if that data is 
used and shared in unintended or unexpected ways (\eg{} shared with 
third-party SDKs for secondary purposes, as we study next).

\para{Access to permissions for secondary purposes.}
Using the pipeline described in \S\ref{sec:static-analysis}, we find that the
inclusion of third-party libraries is common in
mobile browsers: we find at least one advertising (\eg{} Google Ads, Baidu
Mobile Ads), analytics (\eg{} Crashlytics, AppsFlyer) or social
network (\eg{} Facebook, Umeng) library in \oneLib{} of browsers.
The most common SDK across all browsers is
Google Ads (\perGoogleAds{}), which means that developers can potentially
add their own advertisements on top of those that are present in websites
(however we did not find any instance of this behavior at run time).
We also find that apps from Chinese stores rely on libraries that
target Asian markets (\ie{} Baidu, Umeng or Tencent).

As discussed in
\S\ref{sec:static-analysis}, Android's permission model
allows third-party libraries included in Android apps to inherit the
permissions requested by the host app (\ie the browser). 
We now analyze whether permission-protected methods are
requested by third-party libraries embedded in each browser in
order to infer potential secondary purposes. 
We focus on libraries offering advertising,
analytics and social networking services, as they are more likely to leverage
the set of dangerous 
permissions granted to the host app to collect unique identifiers and
behavioral data.

Table~\ref{table:perm_usage} shows, for each permission labeled by
Android as dangerous, the percentage of browsers in which
at least one permission-protected API that requires such permission is present only in
browser code, third-party code or in both.
As we explained in
\S\ref{sec:methodology}, we rely on a mapping from API calls
to AOSP permissions to generate this data.
In addition, when the API call is present only in a third-party SDK
or both in a third-party SDK and first-party code, we specify in
parentheses whether any of the third-party SDK has Advertisement and Tracking
(A\&T) capabilities. Note that due to our manual classification effort,
we might have missed some packages that have A\&T capabilities.
This distinction is important as non-A\&T SDKs might
access permissions for features inherent to the browser
while SDKs with tracking capabilities might use personal data from users
for secondary purposes.

We find significant fractions of third-party SDKs with A\&T capabilities
using dangerous permissions.
The results show that most permissions are requested by A\&T companies for
accessing unique identifiers (\texttt{READ\_PHONE\_STATE} by SDKs such as
Facebook Ads or Google Mobile Services), and location info
(\texttt{ACCESS\_FINE\_LOCATION} by SDKs such as Umeng, Amplitude or AppsFlyer).
Note that static analysis allows us to detect only
potentially privacy-harming behaviors, as it is prone to miss
behaviors due to code obfuscation, reflection or
dynamic code loading.
In the next section, we
investigate how the presence of these third-party SDKs and their access to the
permissions of the host browser translates to data dissemination
to third-party servers at run time.

\begin{table}[t!]
  \centering
  \caption{Percentage of browsers where a given permission is accessed only in
  browser code, in third-party code or in both.}
  \resizebox{\columnwidth}{!}{%
    \begin{tabular}{lrrrr}
   \multirow{2}{*}{\textbf{Permission}}      &
      \multirow{2}{*}{\shortstack[c]{\textbf{Browser}\\\textbf{only}}} &
      \multirow{2}{*}{\shortstack[c]{\textbf{SDK only}\\\textbf{(A\&T)}}} &
      \multirow{2}{*}{\shortstack[c]{\textbf{Browser}\\\textbf{and SDK}}} &
      \multirow{2}{*}{\shortstack[c]{\textbf{\# of}\\\textbf{browsers}}} \\ \\
      \midrule
      \texttt{ACCESS\_COARSE\_LOCATION} & 2\% & 83\% (39\%) & 14\% (125\%) & 113\\
      \texttt{ACCESS\_FINE\_LOCATION} & 2\% & 83\% (37\%) & 13\% (133\%) & 112\\
      \texttt{READ\_PHONE\_STATE} & 25\% & 41\% (73\%) & 32\% (144\%) & 55\\
      \texttt{GET\_ACCOUNTS} & 56\% & 28\% (14\%) & 16\% (25\%) & 25\\
      \texttt{SEND\_SMS} & 0\% & 100\% (0\%) & 0\% (0\%) & 1\\

    \end{tabular}%
    }
  \label{table:perm_usage}
\end{table}

\subsection{Observed PII Exposure}
\label{sec:exposure}

We search the network traffic collected during dynamic testing to identify  
whether it contains PII, and if so, where the PII is sent.
Similar to prior work~\cite{reyes2018won}, we look for sensitive data
(as listed in \S\ref{sec:background})
both in clear text and using popular hashing algorithms (MD5,
SHA-1, SHA-224, SHA-256).
For network traffic, we consider HTTP message bodies, decoded with both the
content-encoding header (\eg{} \texttt{gzip}) and content-type header charset.
Of course, any PII not using these encodings will be missed by our analysis.
In the case of files, we look at the content that is written to or read
from them.
Last, we use data from our \emph{baseline browser} to clean up these results.
Specifically, we consider the data shared by the baseline browser to be  
due to the webpage itself and not the browser that visits them, and 
so we ignore it.
Thus, we are left with data that is generated by the function of the browser
and use this for our analysis.

\para{PII exposure in network traffic.}
\label{sec:pii}
We first analyze network traffic (both encrypted and plaintext) to identify whether 
browsers (or embedded third-party SDKs) exfiltrate sensitive data, and
determine which destinations receive this data. 
We emphasize that a number of these identifiers (\eg{} AdID,
Android ID, MAC Addr.) cannot be accessed using Web APIs; thus, for an
endpoint to be receiving them, the browser must be accessing them directly from Android.

We find that PII exposure is extensive: \leakPIIPercentage{} of browsers disseminate at
least one type of PII. The most common type of PII shared over the
network is the AdID, ``a unique, user-resettable ID for
advertising, provided by Google Play services''~\cite{googleAdID}.
Nevertheless, our results also show apps still collect non-resettable
identifiers contrary to Google's best practices~\cite{identifiers}.
In fact, starting with Android 10 (released in September 2019), Google Play added more
restrictive permission requirements to access non-resettable identifiers 
for apps published in the Google Play Store~\cite{identifiersAndroid10}.
Yet we find that collecting and sharing non-resettable identifiers (\eg{}
the device MAC address and Android ID)
still persists.

\begin{table}[!t]
    \centering
    \caption{Number of browsers across datasets sharing PII (by type) via the network.}
    \resizebox{\linewidth}{!}{%
    \begin{filecontents*}{plots/pii_type_store.csv}
category,play_brow,play_pack,pre_brow,pre_pack,alt_brow,alt_pack,all_brow,all_pack
AD ID,83,83,2,1,13,13,98,94
Location Info.,16,16,0,0,25,23,41,39
MAC$\textsubscript{D}$,2,2,0,0,6,5,8,7
Android ID,1,1,0,0,6,5,7,6
Installed Packages,2,2,0,0,1,1,3,3
MITM Cert.,0,0,0,0,1,1,1,1
\textbf{Total},\textbf{92},\textbf{92},\textbf{2},\textbf{1},\textbf{43},\textbf{41},\textbf{135},\textbf{129}
\end{filecontents*}
\pgfplotstabletypeset[
        every head row/.style={after row=\midrule},
        every last row/.style={before row=\midrule},
        col sep=comma,
        string type,
        columns={category,play_brow,pre_brow,alt_brow,all_brow},
        columns/category/.style={column name=\textbf{PII Type}, column type={l}},
        columns/play_brow/.style={column name=\textbf{Google Play}, column type={r}},
        columns/play_pack/.style={column name=\textbf{P}, column type={r}},
        columns/pre_brow/.style={column name=\textbf{Pre-installed}, column type={r}},
        columns/pre_pack/.style={column name=\textbf{P}, column type={r}},
        columns/alt_brow/.style={column name=\textbf{Alt. Markets}, column type={r}},
        columns/alt_pack/.style={column name=\textbf{P}, column type={r}},
        columns/all_brow/.style={column name=\textbf{All Origins}, column type={r}},
        columns/all_pack/.style={column name=\textbf{P}, column type={r}}
    ]{plots/pii_type_store.csv}}
    \label{table:pii_type_store}
\end{table}

We observe PII dissemination regardless of the origin of the
browser (Table~\ref{table:pii_type_store}). 
The types of PII exposed span all the categories of PII tested, 
even for apps from the Google Play Store (where there are stricter rules 
for compliance with data-collection policies~\cite{playprotect}).
Perhaps in part due to such policies, a larger percentage (\altNonresettableRatio{}) 
of browsers originating from Chinese stores collect
non-resettable IDs when compared to those from Google Play (\playNonresettableRatio{}).
Similarly, a larger percentage of Google Play browsers (\playResettableRatio{}) collect resettable IDs,
as compared to those from Chinese stores (\altResettableRatio{}).

Next, we investigate the destinations that receive this PII, a
complete list of these is in Appendix~\ref{sec:pii_end_host_full}.
As expected, most of these destinations align with our static-analysis
findings(\S\ref{sec:perms}). Google SDKs (which can be used for
tracking and advertisement purposes) such as Google Analytics and 
Firebase are commonly included in mobile
browsers and thus they frequently transmit PII.
Similarly, we observe that well-known social media companies, such as Facebook
and Twitter, also receive PII from browsers.
Our SDK analysis showed that apps from Chinese stores rely more often
on providers specialized in the Chinese market, a finding that is confirmed
by our runtime analysis (\eg{} Tencent and Baidu).

The presence of SDKs in apps does not always
result in PII being shared with these parties (\eg{} Umeng).
On the contrary, dynamic analysis allows us to find third parties
that receive sensitive data (\eg{} Alibaba) but went unnoticed during our
static analysis. This shows the importance of complementing static analysis
with a run-time analysis of browser behavior.
In terms of the type of PII most commonly shared, we find that those related to
targeted advertisement and tracking of users are the most prevalent (namely the
AdID and the device geolocation).
This is not surprising, as third-party SDKs often track users, 
\eg{} to build comprehensive profiles of their behavior and
preferences~\cite{bashir2018diffusion}.
We also find 3 apps \textbf{bridging IDs}, that is, sending 
both resettable and non-resettable identifiers to the same host
(a domain that belongs to UC). This goes against Google's
policy~\cite{identifiers, g_policy} as it allows companies to track users
longitudinally even if they do reset their AdID. Two of these
apps come from Chinese stores and one (UC Browser Turbo) is
available on the Google Play Store.

\subsection{Browsing History Exposure}
\label{ref:history}
To understand the exposure of browsing history, we search for
network requests that contain the names of the automatically visited 
websites. As with our analysis of other types of PII in \S\ref{sec:exposure},
we account for a wide range of 
content and character encodings (\eg{} gzip, JSON, hashed values).
We find that
\totalHistoryLeakBrowsers{} browsers expose the identity of a majority of 
visited websites to an unrelated destination (\ie{} one not contacted when 
loading the tested websites in a baseline browser).
We manually confirm that these domains are not included in any of the websites' source code.
To understand why browsing history is being shared, we manually
analyze the payload of the requests and group behavior into two categories:
those where the browsing history is exposed in support of primary browser functionality 
and those where we could not find any legitimate justification for such behavior.

\begin{table}[!t]
    \centering
    \small
    \caption{Number of browsers sharing visited domains via search and suggestion queries (* First party).}
    \begin{filecontents*}{plots/history_feature_end_host.csv}
company,browsers,packages
Google Suggest,34,33
Google Search,16,16
Yahoo Suggest,2,2
Yandex Favicon*,2,2
Ninesky Favicon*,1,1
Opera Sitecheck*,1,1
Yahoo Search,1,1
Yandex Favicon,1,1
Baidu Suggest,1,1
Opera Sitecheck,1,1
\textbf{Total},\textbf{60},\textbf{59}
\end{filecontents*}
\pgfplotstabletypeset[
        col sep=comma,
        string type,
        columns={company,browsers},
        columns/company/.style={column name=\textbf{Destination Service}, column type={l}},
		columns/packages/.style={column name=\textbf{\# of Packages}, column type={r}},
        columns/browsers/.style={column name=\textbf{\# of Browsers}, column type={r}},
        every head row/.style={after row=\midrule},
        every last row/.style={before row=\midrule},
    ]{plots/history_feature_end_host.csv}
    \label{table:history_feature_end_hosts}
\end{table}

First, we find \historyLeakersFeatures{} browsers where we
identified a feature that requires this data, \ie{} search and
suggestion APIs, site checks, compatibility checks, URL safety checks and
favicon services. 
None of these requests included any other type of PII (such as unique identifiers).
We manually inspected the code of one of the browsers in which we identify such
behaviors and confirm that indeed, whenever the user inputs a domain,
this generates a requests to a search suggestion API (in this
particular case either Google or DuckDuckGo).
This is in line with behavior reported by previous work~\cite{leith2021web}.
Table~\ref{table:history_feature_end_hosts} shows the different
services that we identified in our analysis.
Querying a search API might be expected in some use cases, \eg{} when
the user inputs a term that is not a valid URL.
Nevertheless, when a user enters a complete URL, they may not expect 
or want the URL to be exposed to another party.

\begin{table}[!t]
    \rowcolors{2}{gray!25}{white}
    \centering
    \caption{Number of unique browsers sharing browsing history and PII with
    other parties (* First party), allowing to link the history to a unique user.
  }
    \resizebox{.9\linewidth}{!}{
    \begin{filecontents*}{plots/history_end_host.csv}
company:browsers:packages:android_id:imei:screen_info:location_info:Ad_id:device_mac:wifi_mac:wifi_name:device_ip:installed_packages:os_info:misc
AppsFlyer:8:8::::1:8:::::::
Casale Media:7:7::::::::::::
Rubicon Project:6:6::::::::::::
Yandex*:3:3:::::1:::::::
Moat:3:3::::::::::::
Verizon:2:2:::::1:::::::
360:2:2::::1::::::::
Adnxs:2:2::::::::::::
\textbf{Total}:\textbf{37}:\textbf{37}::::\textbf{4}:\textbf{10}:::::\textbf{1}::
\end{filecontents*}
\pgfplotstabletypeset[
        every head row/.style={
            after row=\midrule
        },
        every last row/.style={
          before row={
            \midrule
            \rowcolor{white}
          },
        },
        col sep=colon,
        string type,
        columns={
            company,
            browsers,
            location_info,
            Ad_id,
            imei,
            device_mac
        },
        columns/location_info/.style={
            column name=\textbf{Loc.},
            column type={r}
        },
        columns/device_mac/.style={
            column name=\textbf{MAC$\textsubscript{D}$},
            column type={r}
        },
        columns/Ad_id/.style={
            column name=\textbf{AdID},
            column type={r}
        },
        columns/imei/.style={
            column name=\textbf{IMEI},
            column type={r}
        },
        columns/wifi_mac/.style={
            column name=\textbf{MAC$\textsubscript{W}$},
            column type={r}
        },
        columns/screen_info/.style={
            column name=\textbf{Screen},
            column type={r}
        },
        columns/company/.style={
            column name=\textbf{Destination},
            column type={l}
        },
		columns/packages/.style={
            column name=\textbf{\# of P},
            column type={r}
        },
        columns/browsers/.style={
            column name=\textbf{\# of Brow.},
            column type={r}
        } 
    ]{plots/history_end_host.csv}
    }
    \label{table:history_end_hosts}
\end{table}

Table~\ref{table:history_end_hosts} shows the destinations that receive
browsing history along with PII for the cases in which could not identify a
feature. Destinations found in only one browser are shown in
Appendix~\ref{sec:history_end_host_extended}.
Of the \historyLeakersNonFeatures{} browsers where we cannot identify a 
feature that requires sharing browsing history, we find that
\historyLeakersWithPII{} browsers send this data 
to endpoints along with unique user identifiers (the AdID in 9 browsers,
location data in 3 browsers and both of these in one browser).
Browsers that send both visited websites and other PII
to destinations can be harmful to user privacy because they have the ability to
link the browsing history to an individual.
We find browsing history exposed to third-party organizations
known for offering tracking and advertising solutions,
such as AppsFlyer~\cite{AppsFlyer}, Firebase~\cite{Firebase}
or Verizon (owner of SDKs such as Flurry~\cite{Flurry}). 

To further understand the privacy risks posed by these third-party SDKs, we
manually analyze some of their code.
For instance, the MiOTA browser implements its own
version of Android's \emph{WebView}.
After every request has finished, the browser runs
JavaScript code that sends the visited URL
along with identifiers (\eg{} the Android ID) to a first-party domain.
Another example is Stealth Browser which sends
the visited URL to a destination that belongs to Fillr~\cite{fillr}.
The browser's description on Google Play~\cite{stealth-browser}
confirms that this browser embeds the Fillr third-party SDK, explaining why we see
these requests in our test. 
Ironically, both browsers claim to offer privacy as one of their features
in their descriptions.
Some of the observed browser-history sharing might serve to improve the browsing 
experience.
For instance, we find destinations related to the browsers' own companies
(for browsers like \emph{UCWeb}, \emph{Opera}, or \emph{Kiddoware}), and
argue that this data collection 
could be used for telemetry, safe browsing or parental control.
Users might decide that the benefits of such collection may offset the privacy risks.

\section{Multidimensional Analysis}
\label{sec:privacy_implications}

 \begin{figure*}[t!]
   \centering
   \includegraphics[width=\linewidth]{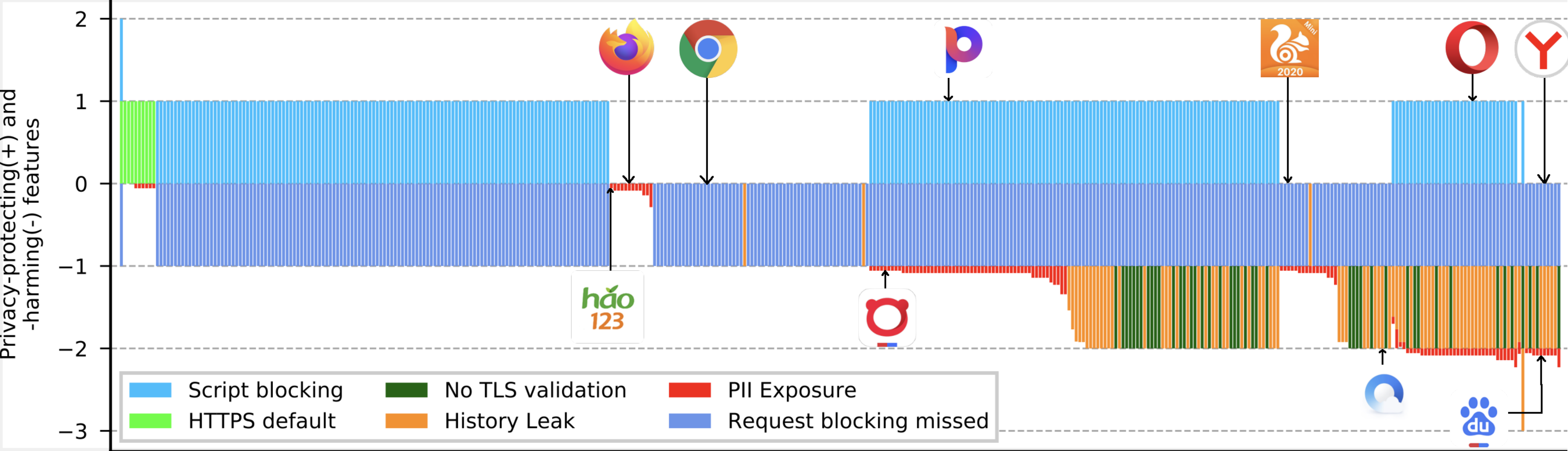}
   \caption{Quantification of browser's privacy-enhancing and -harming behaviors. Browsers icons can be found in Table \ref{table:top_10}. \ml{also: what are the best and the worst (that aren't in the most popular lists?)}}
   \label{fig:browser_features_combined}
 \end{figure*}

 In this section, we combine the findings of previous sections that showed
 instances of privacy enhancing and harming behavior by browsers
 to gain a global picture of the privacy disposition of browsers through a
 multidimensional lens. This allows us to 
 identify instances where browsers are relatively helpful or harmful when it comes 
 to exposing users to data collection, and identify other interesting cases of unexpected 
 behavior.

\para{Methodology.}
To quantify the privacy disposition of each browser, we translate 
each observed behavior into a 
quantified, normalized score. Specifically, we focus on four main 
categories of privacy-impacting behavior for which we assign each browser a
score between 0 and 1.  
$(i)$ \emph{Blocking tracking content or allowing requests to tracking
  services}, for which we use simple binary values.
$(ii)$ \emph{Connection security} with binary values as well. 
$(iii)$ \emph{PII exposure}, for which we assign three levels of values, highest for
\emph{Non-resettable IDs}, medium for \emph{Resettable IDs}, and lowest for
other identifiers. We sum these factors across browsers and normalize them to 
a maximum value of 1.
$(iv)$ \emph{Sharing browsing history}, for we use binary values: if a browser
shares the vast majority 
of websites visited with another party, the value is 1; otherwise it is zero.

\para{Results.}
Figure~\ref{fig:browser_features_combined} represents for each browser
(x-axis), the observed behaviors that improve user privacy in the positive y-axis
and the ones that may  harm them on the negative y-axis. 
Thus, we combine information from all our tests to help understand how 
each browser protects or harms privacy.
The graph is ordered by decreasing privacy protection from left to right, 
meaning the browsers that mostly protect user privacy are on the left 
and those with significant potential harms to user privacy are on the right.

Figure~\ref{fig:browser_features_combined} shows wide variations across browsers
in terms of behaviors 
that impact user privacy, and emphasizes the importance of a multidimensional 
analysis when considering the privacy disposition of a browser.
Of the 5 worst browsers (towards the right end in
Figure~\ref{fig:browser_features_combined}), 4 are available on Google Play (Savannah,
Star, Yandex, and Yandex Beta).
The Yandex browser is extremely popular ($>$ 100 million downloads, but harms 
user privacy by collecting Advertisement IDs (red in
Figure~\ref{fig:browser_features_combined}) along with browsing history (orange
in Figure~\ref{fig:browser_features_combined}).
Meanwhile, a browser found on a Chinese store TxWebLibrary
collects user history while disregarding TLS certificates (dark green in
Figure~\ref{fig:browser_features_combined}).
On the left end, browser that protect user privacy tend to implement HTTPS by
default and block content like trackers and ads (FOSS Browser).
Among the most popular browsers, Chrome and Firefox are ranked best for privacy, 
Phoenix is middle-of-the-pack, and Opera, UC-mini, and Yandex fare poorly. 
We also find interesting cases with mixed impact on privacy between these extremes. 
For example, out of the \easylistBrowsers{} browsers that block tracking content, we see that
\blockAndMiss{} also allow tracking requests, \blockAndPII{} expose PII,
\blockAndShareHistory{} share the browsing history, and \blockAndCertFail{}
fail to validate certificates.
Finally, KKBrowser
(with two different signatures in our dataset) uses 
HTTPS by default and blocks tracking content, but also fails to correctly validate
 TLS certificates and exposes PII.

\section{Related Work}
\label{sec:related_work}
Previous work leveraged static~\cite{arzt2014flowdroid}
and dynamic~\cite{enck2014taintdroid, le2015antmonitor,
razaghpanah2015haystack} analysis to identify security and privacy
threats 
in mobile apps, reporting on: malicious and insecure
behaviors~\cite{lindorfer2014andrubis,
  chen2015massvet,onwuzurike2018droids,reardon201950, scoccia2020leave,
wang2019characterizing, kotzias2021did},
presence of third-party code~\cite{ma2016libradar,razaghpanah2018apps,
shuba2018nomoads, pan2018panoptispy, ren2016recon} and lack of 
transparency when it comes to privacy~\cite{okoyomon2019ridiculousness, zimmeck2016automated}.
Researchers have shown that privacy issues can occur in all types of Android
apps, regardless of their origin~\cite{wang2018beyond, gamba202preinstalled},
or target audience~\cite{valente2017security, reyes2018won}.
Previous work has also shown the pitfalls of static 
techniques~\cite{poeplau2014execute,continella2017obfuscation, faruki2014evaluation,
maiorca2015stealth} and dynamic techniques~\cite{choudhary2015automated,
singh2014automated, textexerciser20,gamba202preinstalled};
finally showing that combining both analysis techniques can be useful for
better understanding security and privacy issues, such as the use of covert and
side channels~\cite{reardon201950} or the privacy risks of analytics
libraries~\cite{liu2019privacy}.
Our work combines static and dynamic 
analysis to provide a comprehensive view of privacy issues in mobile browsers.

In the mobile browser ecosystem, Niu \etal{} highlighted the security
problems of browsers on mobile devices in comparison to those on
desktops~\cite{niu2008iphish}. Leith studied privacy issues of five major mobile browsers~\cite{leith2021web},
showing that indeed some browsers are more privacy-protecting than others. Luo \etal{} studied the evolution of UI vulnerabilities in mobile 
browsers~\cite{luo2017hindsight}, as well as provided a longitudinal analysis
of security mechanisms supported by mobile browsers~\cite{luo2019time}.
Vila \etal{} demonstrated the possibility to perform side-channel attacks on Chrome's event loops to identify websites and user behavior~\cite{vila2017loophole}.
Lin \etal{} explored the privacy and security threats of Chromium-based browsers'
autofill functionality, allowing attackers to extract user data and studying
the potential for such an attack in the wild~\cite{lin2020fill}. Wu \etal{}~\cite{wu2014analyzing} highlighted vulnerabilities in mobile browsers that could give attackers access to users' cookies and browsing history through local file access. Kondracki \etal{} identified potential security issues introduced by data saving
browsers~\cite{kondrackimeddling}.
Furthermore, several related papers identified possible
attacks on Android's WebView~\cite{luo2012touchjacking, luo2011attacks,
chin2013bifocals,webview:neugschwandtner2013,beerbridge,mutchler15:mobilewebapps}, which 
is used by some browsers and many other types of mobile apps to display 
web content. Most recently, Krause~\cite{krause-tiktok} used an instrumented website to reveal that that 
in-app browsers, such as the one in TikTok, collect extensive data about user interactions with websites via JavaScript. However, in-app browsing interfaces are out of scope for our work, as they do not allow users to visit arbitrary URLs as freely as with a dedicated browser. In contrast, we aim to analyze the whole ecosystem of Android
browsers, from those that are more well-known and available in the
Google Play Store to those less-known and from Chinese stores. 

In terms of server-side privacy attacks, Eckersley looked at the possibility
to fingerprint browsers based on their uniqueness~\cite{eckersley2010unique}. 
Vastel \etal{} further showed that browser fingerprints are a viable option to 
perform long-term tracking of users, even when those fingerprints change over
time~\cite{vastel2018fp}.
Das \etal{} showed how mobile specific sensors are used for
tracking~\cite{das2018web} and Marcantoni \etal{} have extended this work,
studying the prevalence of WebAPI sensor accesses on the
web~\cite{marcantoni2019large}.
%
To the best of our knowledge, our work is the first to investigate
\emph{privacy implications} of Android browsers from different
sources, relying on a novel methodology to study whether browsers
promote user privacy by blocking data collection by other parties,
or whether they harm it by collecting and/or sharing user data. 
We develop novel analysis techniques and apply them to a large collection of 
mobile browsers.

\section{Discussion}
\label{sec:discussion}

\para{Limitations.}
We limit our analysis to the Android ecosystem of mobile browsers,
due to the open nature of Android, which permits browsers from
very different origins and business models.
In contrast, Apple's iOS restricts all 
browsers to use WebKit~\cite{appStore}, and users were not allowed to change 
their default
browser until the recent iOS 14~\cite{ios14}. 
At the methodology level, it is important to note that our static analysis
results (\S~\ref{sec:static-analysis})
could be impacted by apps making use of code obfuscation and dynamic code
loading. This is a well known limitation of
static analysis, and it can impact our ability to find
third-party SDKs and identify calls to permission-protected methods.
Our dynamic analysis pipeline (\S~\ref{sec:dynamic_pipeline_analysis})
also presents limitations, preventing us from 
testing \noTest{} browsers dynamically.
Moreover, our behavioral analysis represents a lower bound on potentially problematic 
browser behaviors, as we do not exhaustively cover all code paths due
to limitations of current fuzzing methods.
We also refrain from looking at the purposes of data collection, due to the
difficulty and scalablity issues caused by trying to automatically extract
this information from privacy policies at scale.
The number of pre-installed browsers that we successfully tested is
small due to their native dependencies which cannot be reproduced
on our test device~\cite{gamba202preinstalled}.
Finally, we report on all run-time data collection from mobile browsers by approving 
all permission requests,
which may differ from how any individual user gives consent.

\para{Mitigation.}
Given the lack of transparency into how browsers impact privacy and security, we argue 
that app stores could play a vital role in mitigating harms for users. Specifically, 
app stores can use our technique (building on our open-source code) to measure the privacy 
and security of browsers and decide on whether to publish a browser based on the results. 
We believe this would incur low additional cost since app stores already review apps 
and run automated tools to analyze them, and doing so could protect large numbers of users. 
In addition, stores (or other sites) can make a transparency report available to users at 
install time, similar to the data Apple and the Google Play store report to users regarding an
app's access and use of personal or sensitive data.

\para{Lessons learned.} 
At first glance, a privacy analysis of mobile browsers seems straightforward. However, we learned 
a number of lessons in addressing key challenges that demonstrated it was far from trivial.
One such challenge was dealing with web page dynamics, which make it difficult to reliably 
infer browser behavior. Our web page replay approach, while not entirely novel, required substantial 
effort to properly handle page dynamics---including those that occur even when the page source 
stays the same. Another challenge was determining how a browser modifies page content without 
instrumenting or analyzing the code of hundreds of browsers. We found that injecting our own 
Javascript into webpages provided a surprisingly straightforward way to capture this information 
across all browsers without needing manual instrumentation or analysis. Last, we stress that 
simply looking at PII exposure is not enough when considering the purpose of apps like browsers. 
In our case, we found browsers to share potentially highly sensitive metadata (browsing history) 
with other parties. For future work, it is important to consider such non-standard types of sensitive 
data that can be exposed by special-purpose apps that handle sensitive data beyond those protected 
by OS permissions.

\para{Future work.} Investigating regulatory compliance of consent, 
opt-out, and the accuracy of privacy policy text 
is an interesting area for future work, but it complicated by the complexity and scalability issues of
automatically inspecting privacy policies. 
Interestingly, only \onboardingBrowsers{} of the browsers had some kind of initial on-boarding screen
that the user has to interact with before using the browser, while 
\privacyPolicyFirstBrowsers{} implicitly or explicitly present users with
a privacy policy.
Therefore, for the majority of browsers, we arguably never give explicit consent to data
collection (other than by simply installing and running the browser).
We note that we also looked for file modification by browsers in an attempt to
find apps writing sensitive information to the external storage
as previous work has shown the risks of exposing of PII to external storage,
and how this can be used as a covert channel~\cite{bianchi2017flauth, reardon201950}.
However, we did not find any instance of this behavior during our tests.

\section{Conclusion and key findings}
\label{sec:conclusion}

This paper conducted the first large-scale privacy analysis of Android browsers.
To that end, we analyzed
a set of \totalBrowsers{} browsers collected from different sources, including the Google
Play Store, popular Chinese stores, and pre-installed apps.
We developed a custom, novel methodology that combines static and dynamic
analysis to identify and quantify privacy-protecting and -harming behaviors.
We show that some browsers have the ability to protect user privacy,
as \easylistBrowsersPercentage{} of these browsers block tracking scripts, and
\percTrackerBlockers{} block access to protected JavaScript APIs
(\S~\ref{sec:content_modification} and \S~\ref{sec:blocking_apis}).
However, we also find security issues in \httpsFailPercentage{} browsers that
fail to validate TLS certificates, and discover that only \domainHttpsBrowsersPercentage{}
of browsers default to HTTP\textbf{S} (\S~\ref{sec:connection_security}).
We also find four browsers that modify webpages and inject scripts into
loaded webpages.
Our analysis of mobile browser code shows that these apps request a wide range of 
permissions and that the personal data protected by them can be accessed 
by third-party libraries that may use the data for secondary purposes (\S~\ref{sec:perms}).  
To identify a lower bound of how much personal data is exposed by these browsers 
at runtime, we analyzed network traffic while browsers visit a controlled set of websites.  
We find evidence of \leakPIIPercentage{} of browsers disseminating at
least one type of PII, including resettable and non-resettable identifiers
(\S~\ref{sec:exposure}).
Given the increased flexibility for users to select their own default browser and 
wide range of privacy-impacting behaviors observed, we argue there needs to be 
greater transparency and auditing of mobile browsers, which our techniques  
can readily inform.

\begin{acks}

We thank the reviewers for their valuable feedback on improving our paper. We also thank Jakob Bleier for his assistance with the browser engine attribution. 

This research was partially funded by the NSF under SaTC-1955227. This research also received funding from the Vienna Science and Technology Fund (WWTF) through project ICT19-056, as well as SBA Research (SBA-K1), a COMET Centre within the framework of COMET - Competence Centers for Excellent Technologies Programme and funded by BMK, BMDW, and the federal state of Vienna.
The COMET Programme is managed by FFG.

IMDEA Networks' researchers are funded by EU's H2020 Program 
(TRUST aWARE Project, Grant Agreement
No. 101021377) and the Spanish Ministry of Science (ODIO Project,
PID2019-111429RB-C22). Dr. Narseo Vallina-Rodriguez is funded by a Ramon y
Cajal Fellowship from 
the Spanish Ministry of Science and Innovation.

\end{acks}

\bibliographystyle{ACM-Reference-Format}
\bibliography{refs}

\appendix
\appendix


\begin{table}[h]
  \centering
  \tabcolsep5pt
  \caption{Data collected during dynamic analysis.}%
  \resizebox{\columnwidth}{!}{%
    \begin{tabular}{ll}
    \textbf{Analysis Component (Technique)} &
      \multicolumn{1}{c}{\textbf{Collected Data}}\\
    \midrule
    \director\ (\texttt{adb logcat}) & Logcat dump\\
    \director\ (\texttt{adb screencap}) & Mobile device screenshots\\
    \director\ (\texttt{adb-sync}) & Files on external storage\\
          \director\ (\texttt{adb}) + Mobile Device (\texttt{fsmon}) &
          Filesystem changes on external storage \\ 
    Mobile Device (\texttt{tripwire}) & Browser DOM\\
    \gateway\ (\texttt{mitmproxy}) & TLS traffic\\
    \gateway\ (\texttt{tcpdump}) & All network traffic\\
  \end{tabular}%
  }
\label{table:data_collected}
\end{table}

\section{Website Selection}
\label{appendix:website_selection}
\noindent

\begin{itemize}[leftmargin=*]

\item \para{Honeypage.}
To understand how browsers handle referral links (whether they modify
them to generate revenue by users' purchases), social media plugins, and ads
(whether they block them to enhance users' privacy), we use a honeypage.
This honeypage is inspired by
Tsirantonakis~\etal{}~\cite{tsirantonakis2018large} and contains
\emph{fake advertisements}, an Amazon referral link, and social media plugins.
The source code for this page can be found on Pastebin (\url{https://pastebin.com/qd25cNmC}) and in our artifact repository.

\item \para{Permissions page.}
We extend a permissions test page~\cite{permissionsSite} to add tests for an
exhaustive list of JavaScript WebAPIs.
We also modify this site to send result data to our testbed and use it to test
access to device sensors.

\item \para{Domain without protocol.}
To see if browsers pick \emph{http://} or \emph{https://} by default when a user
visits a domain without specifying a protocol, we use a webpage that supports
both these protocols and omit mentioning the protocol.
The webpage is under our control and performs no redirection but serves content
over the requested protocol.

\item \para{HTTPS webpage.}
  To understand the nature of TLS connections a browser uses, including TLS
  versions, ciphers supported \etc, we use a HTML webpage hosted on a server
  that supports HTTPS.

\item \para{Popular webpages.}
 To understand how browsers handle regular websites, we test the 16 websites listed in Table~\ref{table:alexa_list}, one from each category of Alexa's category
  rankings~\cite{alexaCategory}. (Note that this feature was retired in September 2020.)

\begin{table}[h!]
    \centering
    \small
    \caption{Popular websites from Alexa's category rankings.}
    \begin{filecontents*}{plots/alexa_list.csv}
website,category
https://www.xvideos.com/,Adult
https://www.office.com/,Business
https://www.yahoo.com/,Regional
https://m.twitch.tv/,Games
https://www.nih.gov/,Health 
https://www.patreon.com/,Society
https://finance.yahoo.com/, Home
https://www.researchgate.net/,Science
https://www.reddit.com/,News
https://www.booking.com/,Recreation
https://stackoverflow.com/,Reference
https://www.amazon.com/,Shopping
https://www.espn.com/,Sports
\end{filecontents*}
\pgfplotstabletypeset[
        col sep=comma,
        string type,
        columns={category,website},
        columns/website/.style={column name=\textbf{Website}, column type={r}},
        columns/category/.style={column name=\textbf{Category}, column type={l}},
        every head row/.style={after row=\midrule},
    ]{plots/alexa_list.csv}
    \label{table:alexa_list}
\end{table}

\end{itemize}

\begin{table}[!t]
    \rowcolors{2}{gray!25}{white}
    \centering
    \caption{Entities receiving PII from browsers (* First party).}
    \resizebox{0.8\linewidth}{!}{%
    \begin{filecontents*}{plots/pii_hosts_full.csv}
host,installed_packages,device_ip,location_info,android_id,mitm_cert,device_mac,Ad_id,total
Firebase,,,,,,,33,33
Facebook,,,2,,,,24,24
Alibaba,,,14,1,,2,1,18
AppsFlyer,,,1,,,,13,13
Verizon,,,,,,,9,9
Baidu,,,7,,,,,7
Adjust,,,,,,,7,7
Twitter,,,,,,,6,6
umsns,,,,6,,1,1,6
Appnext,,,,,,,6,6
OneSignal,,,,,,,5,5
StartAppService,,2,1,,,,4,4
Google Analytics,,,1,,,,3,4
AppsGeyser,,,,,,,3,3
StartApp,,,,,,,3,3
ctobsnssdk,,,,,,3,1,3
toutiao,,1,2,,,,,3
Mobvista,,,1,,,,1,2
Microsoft,,,2,,,,,2
360,,,2,,,,,2
Taboola,,,,,,,2,2
smardroid,,,,,,,1,1
Google devicecheck,,,1,,,,,1
Yahoo Search,,,1,,,,,1
Cloudflare Certcheck,,,,,1,,,1
Tencent,,,1,,,,,1
Mindworks,,,,,,,1,1
Urora,1,1,,,,,,1
impression,,,,,,,1,1
Brave,,,1,,,,,1
Opera*,,,,,,,1,1
Amap,,,1,,,,,1
Kuaishou,,,1,,,,,1
yladm,,,,,,1,,1
1look,,,,,,1,,1
AdColony,,,,,,,1,1
UCWeb*,,,1,,,,,1
uodoo,,,1,,,,,1
UCWeb,,,,,,1,1,1
yohoads,,,,,,,1,1
Flymobi,,,,,,,1,1
suibyuming,,,1,,,,1,1
oakmastering,,,,,,,1,1
cloudmobi,,,,,,,1,1
MiOTA*,,,1,,,,,1
Amazon,,,,,,,1,1
Yandex*,,,,,,,1,1
Mail.ru*,1,,,,,,1,1
Amplitude,,,,,,,1,1
Baidu*,,,1,,,,,1
tclclouds,,,,,,1,,1
Xiaomi,,,1,,,,,1
Opera,,,,,,,1,1
Vserv,1,,,,,,1,1
Vmax,,,,,,,1,1
duapps,,,,,,,1,1
\textbf{Total},\textbf{3},\textbf{4},\textbf{41},\textbf{7},\textbf{1},\textbf{8},\textbf{98},\textbf{135}
\end{filecontents*}
\pgfplotstabletypeset[
        col sep=comma,
        string type,
        columns={host,Ad_id,location_info,device_mac,total},
        columns/host/.style={column name=\textbf{Destination}, column type={l}},
        columns/location_info/.style={column name=\textbf{Loc}, column type={r}},
        columns/device_mac/.style={column name=\textbf{MAC$\textsubscript{D}$}, column type={r}},
        columns/Ad_id/.style={column name=\textbf{AdID}, column type={r}},
        columns/android_id/.style={column name=\textbf{Android ID}, column type={r}},
        columns/imei/.style={column name=\textbf{IMEI}, column type={r}},
        columns/wifi_name/.style={column name=\textbf{WiFi}, column type={r}},
        columns/wifi_mac/.style={column name=\textbf{MAC$\textsubscript{W}$}, column type={r}},
        columns/device_ip/.style={column name=\textbf{IP}, column type={r}},
        columns/contact_info/.style={column name=\textbf{Contact Info}, column type={r}},
        columns/screen_info/.style={column name=\textbf{Screen Info}, column type={r}},
        columns/total/.style={column name=\textbf{Total}, column type={r}},
        every head row/.style={after row=\midrule},
        every last row/.style={before row=\midrule},
    ]{plots/pii_hosts_full.csv}}
    \label{table:pii_hosts_full}
    
\end{table}

\section{Browsers sharing PII}
\label{sec:pii_end_host_full}
Table~\ref{table:pii_hosts_full} lists all the observed destinations (manually grouped by the
organization that owns them), as well as the type of data that they receive.

\section{Browsing History Exposure}
\label{sec:history_end_host_extended}

The remaining rows of Table~\ref{table:history_end_hosts}, listing destinations
that receive browsing history along with PII (if any).

\begin{table}[h]
    \rowcolors{2}{gray!25}{white}
    \centering
    \caption{
    Number of unique browsers sharing browsing history and PII
      together, allowing other parties to link the history to a unique user
      (* First party).}
    \begin{filecontents*}{plots/history_end_host_extended.csv}
company:browsers:packages:android_id:imei:screen_info:location_info:Ad_id:device_mac:wifi_mac:wifi_name:device_ip:installed_packages:os_info:misc
Oupeng:1:1::::::::::::
Baidu:1:1::::1::::::::
ilovegame:1:1::::::::::::
baoruan:1:1::::::::::::
FDVR*:1:1::::::::::::
Google Datamixer:1:1::::::::::::
Kiddoware*:1:1::::::::::::
Maxthon*:1:1::::::::::::
Orbitum*:1:1::::::::::::
Fillr:1:1::::::::::::
Tencent*:1:1::::::::::::
fddm:1:1::::::::::::
MiOTA*:1:1::::1::::::::
Bing:1:1::::::::::::
Mail.ru*:1:1:::::1:::::1::
Aonbrowser:1:1::::::::::::
Google Analytics:1:1::::::::::::
Orbitum:1:1::::::::::::
\textbf{Total}:\textbf{37}:\textbf{37}::::\textbf{4}:\textbf{10}:::::\textbf{1}::
\end{filecontents*}
\pgfplotstabletypeset[
        every head row/.style={
            after row=\midrule
        },
        every last row/.style={
          before row={
            \midrule
            \rowcolor{white}
          },
        },
        col sep=colon,
        string type,
        columns={
            company,
            browsers,
            location_info,
            Ad_id
        },
        columns/location_info/.style={
            column name=\textbf{Location},
            column type={r}
        },
        columns/device_mac/.style={
            column name=\textbf{MAC$\textsubscript{D}$},
            column type={r}
        },
        columns/Ad_id/.style={
            column name=\textbf{AdID},
            column type={r}
        },
        columns/imei/.style={
            column name=\textbf{IMEI},
            column type={r}
        },
        columns/wifi_mac/.style={
            column name=\textbf{MAC$\textsubscript{W}$},
            column type={r}
        },
        columns/screen_info/.style={
            column name=\textbf{Screen},
            column type={r}
        },
        columns/company/.style={
            column name=\textbf{Destination},
            column type={l}
        },
		columns/packages/.style={
            column name=\textbf{\# of P},
            column type={r}
        },
        columns/browsers/.style={
            column name=\textbf{\# of Browsers},
            column type={r}
        } 
    ]{plots/history_end_host_extended.csv}
    \label{table:history_end_hosts_extended}
\end{table}

\end{document}